\documentclass[12pt,letterpaper]{article}
\usepackage{amssymb,latexsym,amsmath,amscd,amsthm}
\usepackage{amsfonts}
\usepackage{epsfig}
\usepackage[active]{srcltx}
\usepackage{color}
\usepackage[latin1]{inputenc}
\usepackage[round,authoryear]{natbib}
\usepackage{xspace}
\usepackage{appendix}
\usepackage{placeins}
\usepackage{rotating}
\usepackage{lscape}
\usepackage{caption}
\usepackage{changes}
\usepackage{float}
\usepackage{hyperref}

\usepackage{lipsum}

\renewcommand{\section}[1]{%
\bigskip
\begin{center}
\begin{Large}
\normalfont\scshape #1
\medskip
\end{Large}
\end{center}}

\renewcommand{\subsection}[1]{%
\bigskip
\begin{center}
\begin{large}
\normalfont\itshape #1
\end{large}
\end{center}}

\renewcommand{\subsubsection}[1]{%
\vspace{2ex}
\noindent
\textit{#1.}---}

\renewcommand{\tableofcontents}{}

\bibpunct{(}{)}{;}{a}{}{,}  % this is a citation format command for natbib

\newcommand{\RR}{\mathbb{R}}

\newcommand{\saq}{\texttt{SAQ}\xspace}
\newcommand{\eri}{\texttt{Erik+2}\xspace}

\newcommand{\nj}{\texttt{NJ}\xspace}
\newcommand{\ml}{\texttt{ML}\xspace}
\newcommand{\mll}{\texttt{ML}}

\definecolor{pink}{rgb}{1,0,1}

\newif\ifprivate
\privatetrue

\def\???{\ifprivate {\bf {???}} \marginpar{{\Huge {\bf ?}}}
\else \fi}

\newtheorem{defn0}{Definition}[section]
\newtheorem{prop0}[defn0]{Proposition}
\newtheorem{thm0}[defn0]{Theorem}
\newtheorem{lemma0}[section]{Lemma}
\newtheorem{corollary0}[defn0]{Corollary}
\newtheorem{example0}[defn0]{Example}
\newtheorem{conjecture0}[defn0]{Conjecture}
\newtheorem{notation0}[defn0]{Notation}
\newtheorem{remark0}[defn0]{Remark}
\newtheorem{assumption0}[defn0]{$d$-claw tree hypothesis}
\newtheorem{problem0}[defn0]{Problem}

\newenvironment{lema}{\begin{lemma0}}{\end{lemma0}}

\begin{document}

%\begin{flushright}
%Version dated: \today
%\end{flushright}
\bigskip
%\noindent Invariant versus classical quartet inference
% put in your own RH (running head)
% for POVs the RH is always POINT OF VIEW

\bigskip
\medskip
\begin{center}

% Insert your title:
\noindent{\Large \bf SAQ: semi-algebraic quartet reconstruction method}
\bigskip

% We don't use a special title page; the author information is entered
% like any other text.

% FOOTNOTES: We don't allow them in the manuscript, except in
% tables. Don't include any footnotes in the text.

\noindent {\normalsize \sc Marta Casanellas, Jes\'{u}s Fern\'{a}ndez-S\'{a}nchez$^1$ and Marina Garrote-L\'{o}pez, $^1$}\\
\noindent {\small \it
$^1$Dpt. Matem\`{a}tiques, Universitat Polit\`{e}cnica de Catalunya, Barcelona, Spain}\\
\end{center}
\medskip
%\noindent{\bf Corresponding author:}   \\
Universitat Polit\`{e}cnica de Catalunya, Av. Diagonal 647, 08028-Barcelona, Spain \\\\
%\noindent{E-mail: Corresponding author:}  
%\blfootnote{} 
\let\thefootnote\relax\footnotetext{E-mail addresses: \href{mailto:marta.casanellas@upc.edu}{marta.casanellas@upc.edu} (M. Casanellas), \href{mailto:jesus.fernandez.sanchez@upc.edu}{jesus.fernandez.sanchez@upc.edu} (J. Fern\'{a}ndez-S\'{a}nchez), \href{mailto:marina.garrote@upc.edu}{marina.garrote@upc.edu} (M. Garrote-López)}

% Of course the specific format of addresses may vary according to
% country or other factors. Also, that was just an example email format.
%It's acceptable to add email addresses for authors in addition to the
%corresponding author. These would be placed after "Country."

%\vspace{1in}

\subsubsection{Abstract}
We present the phylogenetic quartet reconstruction method SAQ (Semi-algebraic quartet reconstruction). SAQ is consistent with the most general Markov model of nucleotide substitution and, in particular, it allows for rate heterogeneity across lineages. Based on the algebraic and semi-algebraic description of distributions that arise from the general Markov model  on a quartet, the method outputs normalized weights for the three trivalent quartets (which can be used as input of quartet-base methods). We show that SAQ is a highly competitive method that outperforms most of the well known reconstruction methods on data simulated under the general Markov model on 4-taxon trees. Moreover, it also achieves a high performance on data that violates the underlying assumptions.
%In this paper we present \eri, a topology reconstruction method based on an idea of Eriksson on using phylogenetic invariants. \eri is consistent with the most general Markov model of nucleotide substitution and also allows data coming from mixtures on the same tree topology. \eri provides a system of weights that can be used as input of quartet-based methods. %

\noindent[\textbf{Keywords}: pyhlogenetic reconstruction, general Markov model, quartet inference, algebraic phylogenetics]\\
\vspace{1.5in}

%INTRODUCTION
\newpage
\section{Introduction}
% !TEX root = ms.tex

Phylogenetic reconstruction methods based on algebraic tools have been appointed as potential successful methods for gene reconstruction \citep{allmankubatkorhodes,kubatkochifman2019,chifmankubatko2015,casfer2016}. These methods have the advantage of dealing with unrestricted underlying substitution models that allow for rate heterogeneity across lineages. However, they seem to require a large amount of data. For example, the method \eri \citep{casfer2016} was proven to achieve a high performance under the general Markov model of nucleotide substitution (and allowed also for rate heterogeneity among sites), but needed about 10 000 sites with four taxa to clearly outperform neighbor-joining (NJ) or maximum likelihood (ML). A positive outcome of the algebraic methods for topology reconstruction is that they do not need to estimate the continuous parameters of the underlying evolutionary model. Although they are not ready to be used directly on large trees yet, they could be an excellent input for quartet-based reconstruction methods \citep{Ranwez2001} if a high performance for quartets could be achieved.

In \citet{casfergarrote2020} we proved that algebraic tools may not be enough when dealing with confusing data (e.g. short alignments or quartets with a short internal branch) and that the stochastic nature of the data must be also considered. Theoretically, this could be done via the semi-algebraic description of the general Markov model presented in \citet{AllmanSemialg}. Nevertheless, in the same way that algebraic conditions do not directly use phylogenetic invariants (see \citet{AllmanRhodeschapter4} for an introduction to phylogenetic invariants), the semi-algebraic conditions cannot be used straightforward and separately from the algebraic ones. Providing a quartet reconstruction method that combines both and achieves a high performance is the goal of this article.

We present the method \saq, which stands for Semi-Algebraic Quartet reconstruction method. It is a phylogenetic reconstruction method for DNA alignments on four taxa which assumes a general Markov model of nucleotide substitution (GM henceforth) and which is based on the algebraic and semi-algebraic conditions that characterize data from this model. The underlying model is therefore the most general model of nucleotide substitution on independently and identically distributed sites (unrestricted distribution at the root, unrestricted transition matrices and rate heterogeneity across lineages). The method also outputs normalized quartet weights that can be used as input of quartet-based methods. Note that \saq only aims to reconstruct the topology of the tree, not the substitution parameters.

 We have tested \saq on simulated data under different settings: on a ``tree space" of quartets, on quartets with random branch lengths, and on alignments that are not identically distributed across sites (``mixture data" that violates the assumptions underlying \saq). The data have been generated both under the GM model and a (homogeneous across lineages and) time-reversible continuous-time process (homGTR). Our results show that the method is highly successful, even with short alignments and with data that violates the assumptions of the method. The weights ouput by SAQ are shown to be unbiased and statistically consistent. We also provide a comparison of this method against existing methods such as ML, NJ and \eri and show that \saq largely outperforms all of them for GM data and has a compatible performance for GTR data. Moreover, the results on mixtures of distributions on the same quartet show that the method is also able to deal with heterogeneity across lineages, as it surpasses methods that are specially designed for these data.

\newpage
\FloatBarrier
\section{Methods}
% !TEX root = ms.tex

\subsection{Description of the method}

The method of phylogenetic reconstruction proposed in this paper is  based on the theoretical result by \citet{AllmanSemialg} that we briefly explain here. Given four taxa denoted by $\{1,2,3,4\}$ we consider a DNA alignment of length $N$ and collect the observed relative frequencies of site patterns as a vector in $\RR^{256}$. The coordinates of a vector $p \in \RR^{256}$ are labelled by $x_1x_2x_3x_4$, where each $x_i\in \{A,C,G,T\}$ stands for the observation at leaf $i$.
%as coordinates $p_{x_1x_2x_3x_4}$ where each $x_i\in \{A,C,G,T\}$ stands for the observation at leaf $i$.

The set of the three fully-resolved (unrooted) quartet trees on the set of taxa is denoted as $\mathcal{\tau}=\{12|34, 13|24, 14|23\}$. Given a bipartition $ij|kl$ on the set of taxa and a vector  $p \in \RR^{256}$, we denote by $F_{ij|kl}$ the \emph{flattening} of $p$ according to that bipartition: $F_{ij|kl}$ is the $16\times 16$ matrix whose $(x_ix_j,x_kx_l)$ entry   is the coordinate of $p$ that matches $x_ix_jx_kx_l$ in the convenient order (e.g. the $(AC,GT)$ entry of $F_{12|34}$ is $p_{ACGT}$ while the same entry in $F_{13|24}$ is $p_{AGCT}$).

Briefly speaking, the main result of \citet{AllmanSemialg} states that a distribution $p \in \RR^{256}$ comes from a general Markov (briefly GM) process on the quartet tree $T=12|34$ if and only if the following three conditions are satisfied:
\begin{itemize}
\item[(a)] the marginalization of $p$ over each leaf comes from a Markov process on a tripod tree,
\item[(b)] the matrix $F_{12|34}$ has rank four (or less than four for special parameters),
\item[(c)] after applying sixteen ``$12|34$ leaf-transformations'' to $p$,   the flattening matrices $F_{13|24}(\tilde{p})$ and $F_{14|23}(\tilde{p})$ associated to each transformed vector $\tilde{p}$ are symmetric and positive definite (or positive semidefinite for special parameters).
\end{itemize}
%The method we describe here is based on studying the stochastisity of the parameters of the model describing a certain evolutionary process.

The first condition is independent of the tree topology, so it is not useful for recovering the tree topology. The second condition relies already on the tree topology and has been used in different phylogenetic reconstruction methods \citep[see]{allmankubatkorhodes,kubatkochifman2019,chifmankubatko2015,casfer2016}.
However, this condition is  satisfied by distributions arising not only from trees but also from certain networks (see \cite{CasFerBirkhauser}) and, when restricted to trees, it does not reflect the stochasticity conditions of the transition matrices. The third condition reflects the stochastic nature of the transition matrix at the interior edge.
%The $12|34$ leaf transformations mentioned above perform transformations so that if $p$ had arisen from a Markov process on a tree $12|34$, then these transformed vectors $\tilde{p}$ would have also arisen on $12|34$ but with different transition matrices at the leaves (actually, the same matrices at leaves 1 and 2 and the same at leaves 3 and 4).
%
For each quartet tree $T \in \mathcal{\tau}$, there are sixteen ``$T$ leaf-transformations'' that can be applied to any vector $p\in \mathbb{R}^ {256}$.
If $p$ had arisen from a Markov process on a tree $T'$ (not necessarily the same as $T$) with certain transition matrices, the resulting transformed vectors $\tilde{p}$ would have also arisen on $T'$ but (in general) with different transition matrices. Although these transformations are fully explained in \citep[see]{AllmanSemialg}, we briefly illustrate them in the Appendix.%  (although the $\tilde{p}$ )

According to the results by \citet{casfergarrote2020}, it is important to combine conditions (b) and (c) in  order to obtain successful reconstruction methods for data that might be misleading (that is, small samples or data coming from trees with a short interior edge).

In what follows, we explain how \saq combines conditions (b) and (c). % and makes use of (a) passing by. We explain it in what follows.
For a square matrix $M$, we denote by $psd(M)$ its closest positive semidefinite matrix (see \cite{nearestPSD}) and by $\delta_4(M)$ its distance to the set of rank $\leq 4$ matrices \cite{Demmel}. The rank of $psd(M)$ is smaller than or equal to the rank of  $M$ (see \cite{CFG_linalg}).  For each quartet tree $T$ (in the set $12|34$, $13|24$, $14|23$), given a distribution $p\in \RR^{256}$,  \saq computes a score $s_T(p)$ as follows. If $T=12|34$, we consider the sixteen $12|34$ leaf-transformations $\tilde{p}_i$, $i=1,\dots,16$ of $p$ mentioned in $(c)$ and for each of these vectors we compute
\begin{eqnarray*}
s^i_T:=\frac{\min \left \{\delta_4(psd(F_{13|24}(\tilde{p}_i))),\delta_4(psd(F_{14|23}(\tilde{p}_i)))\right \}}{\delta_4 \left (psd(F_{12|34}(\tilde{p}_i)) \right)}.
\end{eqnarray*}
Then, define $s_T(p)$ as the average of these sixteen quantities. If $T$ is any of the other two trees, $s_T(p)$ is computed analogously  by permuting the roles of the leaves accordingly. Finally \saq outputs the normalized three scores, that is, if $s:=s_{12|34}(p)+s_{13|24}(p)+s_{14|23}(p)$, then
%$$\saq(p):=\frac{1}{s_{12|34}(p)+s_{13|24}(p)+s_{14|23}(p)}\left (s_{12|34}(p),s_{13|24}(p),s_{14|23}(p) \right ).$$
$$\saq(p):=\frac{1}{s}\left (s_{12|34}(p),s_{13|24}(p),s_{14|23}(p) \right ).$$

% \footnotesize
% $$\saq(p):=\left (\frac{s_{12|34}(p)}{s_{12|34}(p)+s_{13|24}(p)+s_{14|23}(p)},\frac{s_{13|24}(p)}{s_{12|34}(p)+s_{13|24}(p)+s_{14|23}(p)},\frac{s_{14|23}(p)}{s_{12|34}(p)+s_{13|24}(p)+s_{14|23}(p)} \right).$$
% \normalsize

If $p$ arises as a distribution on the tree $12|34$ with stochastic parameters, then $F_{12|34}(\tilde{p}_i)$ has rank $\leq 4$ (by (b)) and $psd(F_{12|34}(\tilde{p}_i))$ has also rank $\leq 4$ , so $\delta_4(psd(F_{12|34}(\tilde{p}_i)))=0$. Moreover, if $p$ comes from generic parameters (namely, invertible transition matrices with positive entries and postive distribution at the root node), then $F_{13|24}(\tilde{p}_i)$ (resp.  $F_{14|23}(\tilde{p}_i)$) is a symmetric positive definite matrix by (c) and has rank 16 \citep[][proof of Proposition 5.6]{AllmanSemialg}, so actually $\delta_4(psd(F_{13|24}(\tilde{p}_i)))=\delta_4(F_{13|24}(\tilde{p}_i))>0$  (analogously for $F_{14|23}(\tilde{p}_i)$).  Therefore, $s_{12|34}(q) \rightarrow \infty$ when $q$ approaches a distribution $p$ that was generated  on the tree $12|34$ (with generic stochastic parameters).
%This implies that $s_{12|34}(p)<0$ if $p$ arises as a distribution on the tree $12|34$ with generic stochastic parameters.

On the other hand, $s_{13|24}(q)$ and $s_{14|23}(q)$ tend to zero when $q$ approaches a distribution $p$ generated on $12|34$ with generic stochastic parameters. Indeed, in order to compute $s_{13|24}(p)$ we consider the sixteen $13|24$ leaf-transformations of $p$, which will be denoted as $\hat{p}_i$; then $F_{12|34}(\hat{p}_i)$ has still rank $\leq 4$ (as mentioned above, $\hat{p}_i$ still arises from $12|34$) and its closest positive definite matrix must have rank $\leq 4$ \citep{CFG_linalg}. Thus, $\min\{\delta_4(psd(F_{12|34}(\hat{p}_i))),\delta_4(psd(F_{14|23}(\hat{p}_i)))\}=0$.
% and $s_{13|24}(p)\geq 0$.
Moreover, if $p$ arises from generic parameters, $F_{13|24}(\hat{p}_i)$ has rank 16, is not symmetric anymore and its closest positive definite matrix has generically rank strictly larger than four (see a justification in the Appendix). Hence, in this case $\delta_4(psd(F_{13|24}(\hat{p}_i)))>0$ and $s_{13|24}(p)$ is zero if $p$ comes from generic parameters on the tree $12|34$. A similar argument applies to $14|23$.%see that $s_{14|23}(p)$ is also zero.
By normalizing the scores we get that, if $q\in  \RR^{256}$ is a distribution that tends to a distribution $p$ generated on the tree $12|34$ with generic stochastic parameters, then
$$ % \stackrel{\longrightarrow}{q \rightarrow p}
 \lim_{q \rightarrow p} \saq(q) = \saq(p)=(1,0,0).$$
%The information contained in an alignment of nucleotide sequences is collected in a vector of $P$ whose coordinates are the observed relative frequencies of possible patterns at the leaves. One can perform some transformations on this vector such that the new vector $\tilde{P}$ corresponds to a tree with certain symmetries on it (see Lemma \ref{lema:transf} of the Appendix).
%
%Therefore, if $p$ arises as a distribution of on the tree $12|34$ with stochastic parameters, then we have
%$$\saq(p)=(s_{12|34}(p)\leq 0,s_{13|24}(p)\geq0,s_{14|23}(p)\geq0)$$
%and if the parameters are generic, then $\saq(p)=(<0,>0,>0)$.
According to the \saq method, the correct topology for a distribution $p$ is the topology $T$ for which $s_T(p)$ attains the maximum value. Hereby,  \saq seeks to minimize the average distance of the flattening $F_T$ of the transformations to rank four matrices and to maximize the average distance of the other two flattenings to rank four, under the assumption that these should be positive definite. SAQ can be understood as a quartet inference measure in the sense of \cite{sumner2017}.

In practice, the implementation of an algorithm that computes these scores requires dealing with some technical difficulties. For example, the so-called leaf transformations might require computing the inverse of ill-conditioned matrices derived from the marginalization over two taxa. Moreover, when $p$ is an empirical distribution, then its leaf transformations are not distributions any more. The implementation we provide in
%
%\begin{center} github/...
%\end{center}
\begin{eqnarray*}\label{github_SAQ}
\qquad \mbox{\url{https://github.com/marinagarrote/SAQ-method}.}\vspace{0.2cm}
\end{eqnarray*}
%
%\blue{MArina: pagina del codi, hi sera? Quina parameters se li podran passar?} \red{S?! El posar? a github. L'?nic par?metre que crec que podria variar en el SAq ?s el fildre que determina el m?nim valor que acceptem a les matrius de flattening de les transformacions. Est? fixat a -1 per?podria ser un par?metre. Creieu que cal posar-ho?}
excludes leaf transformations that are far from being distributions (this is the parameter "filter" that can be modified by the user and that could be adapted to the alignment length). %and avoids ill conditioned matrices (parameter "cond\_num").
 %It is in this sense that we claimed that \saq also makes use of (a). \red{aix? ara ho hauriem de treure?}
%\subsection{\eri method}

%\subsection{Maximum likelihood}

%\subsection{Neighbor-joining}

\subsection{Description of the simulated data}

In order to test the new method \saq and compare its performance to other reconstruction methods, we have used the data of  \citet{casfer2016}. These data had been generated under two different models, the general Markov model (GM) and a time-homogenous GTR model (homGTR), under a wide range of systems of branch lengths on quartet trees. We briefly describe these sets of simulated data and refer the reader to \citet{casfer2016} for more details. Branch lengths here are measured as the expected  number of elapsed substitutions per site along the evolutionary process associated to the branch.

\subsubsection{Models of nucleotide substitution}
In reference to the general Markov model (GM), the Markov process on the tree starts with a random distribution and evolves according to random Markov matrices at the edges that correspond to the given branch length. These data had been generated using the software \texttt{GenNon-h} \citep{GenNonh}.

By a time-homogeneous GTR model (homGTR) we mean a continuous-time GTR model that shares the same instantaneous mutation rate matrix $Q$ at all branches of the tree. These data had been generated using \texttt{Seq-gen} \citep{Rambaut1997}.%, the details can be found in \citep{casfer2016}.

%\paragraph{Simulations under the GM model} By using the  software%
%\texttt{GenNon-h} \citep{GenNonh}.
%Given the tree topology of Fig. \ref{tree} (rooted at the parent node of leaves 3 and 4) and values $a,b,c$ for the branch lengths (understood as the expected number of substitutions per site),  a random distribution of nucleotides at the root is generated, as well as  random substitution matrices with the expected amount of substitutions per site. From these, an alignment of four sequences is obtained  according to the corresponding Markov process on the tree \citep{CasKedz}.
%\red{La c no cal...i no se si cal posar tot aixo}

%\paragraph{Simulations under a time-homogeneous GTR model}
%
%In order to generate data evolving under a homGTR model, \texttt{Seq-gen} \citep{Rambaut1997} was used.
%Uniform equilibrium distribution was taken, and the rate matrix underlying \texttt{Seq-gen} alignments on the tree space used in Fig. ???  had rates 2 (A$\rightarrow$C), 7 (A$\rightarrow$G), 4 (A$\rightarrow$T), 3 (C$\rightarrow$G), 1 (C$\rightarrow$T), 5 (G$\rightarrow$T).
% The rate matrix underlying GTR+Gamma-rates had rates 2 (A$\rightarrow$C), 5 (A$\rightarrow$G), 3 (A$\rightarrow$T),
% 4 (C$\rightarrow$G), 1 (C$\rightarrow$T), 2 (G$\rightarrow$T) and the sites were varied according to a gamma distribution
% with parameter $\alpha=\beta$ in the range \replaced{$[0.1,2]$}{$(0,2]$} varying in steps of $0.1$. Small values of this parameter indicate a lot of variation
% across sites \citep{yang1993}.

 \begin{figure}[h]
	\begin{center}
		\includegraphics[scale=0.5]{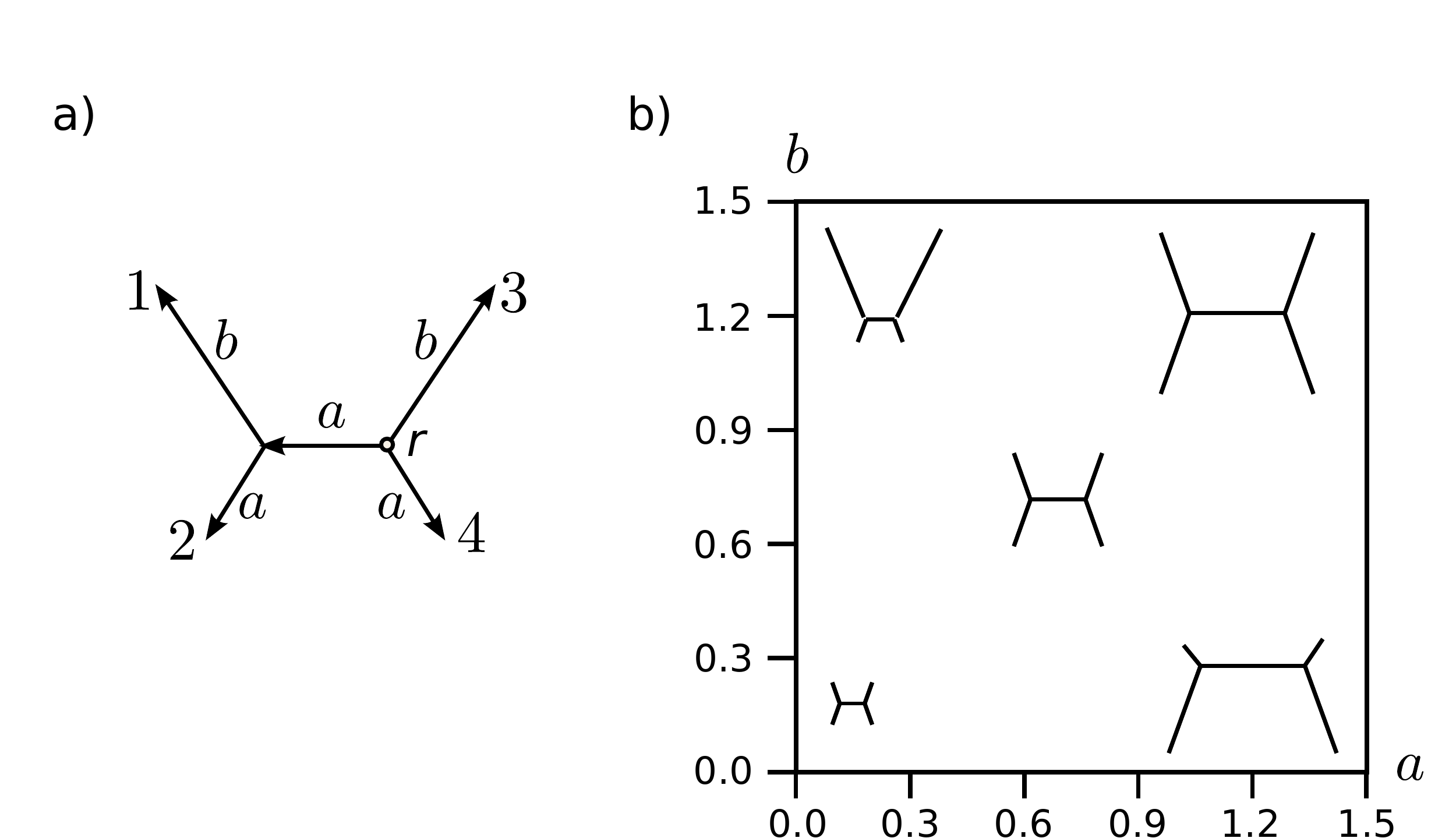}
	\end{center}
	%\vspace{3mm}
	\caption{\label{tree} \footnotesize
		a) 4-leaf tree where the length of two opposite branches and the interior branch are
		represented by $a$; the other two peripheral branches have length $b$; is denoted by $c$.
		%The root $r$ is located at the parent node of leaves 3 and 4.
		Branch lengths will be measured in the expected number of substitutions per site. b) Tree space obtained form the tree in a) when the branch lengths $a$ and $b$ are varied from 0.01 to 1.5 in steps of 0.02.
	}
\end{figure}
\subsubsection{Tree space} We use the parameter space suggested in \citet{huelsenbeck1995} to test
different methods of phylogenetic reconstruction. More precisely, we evaluate the method on a \textit{tree space} (see Figure \ref{tree}.b) where the quartets are as in Figure \ref{tree}.a, and the branch lengths $a$ and $b$ vary between 0 and 1.5 in steps of 0.02. Note that on the top left part of this tree space we find the ``Felsenstein zone'', which contains trees sbuject to the long branch attraction phenomenon.
For each pair $(a,b)$, we have one hundred alignments generated under this setting for each of the two models of nucleotide substitution considered, GM and hom-GTR.

%\subsubsection{Felsenstein zone}
%\red{ho he eliminat}
%
%We consider trees subject to long-branch  attraction, also known as trees in the Felsenstein zone. To this end, on the tree of Figure \ref{tree}.a
%we fix $a= 0.05$, $b=0.75,$ and let the internal branch length $c$ vary in the range $[0.01,0.4]$ (or either we fix $c=0.05$ and call it the \textit{Felsenstein tree}).
%
%For each set of branch lengths, we generated one hundred alignments of different lengths.

\subsubsection{Random trees}
A total of 10 000 alignments are considered, obtained from 4-taxa trees with random branch lengths uniformly distributed either in the interval (0,1) or in (0,3), and generated according to one of the two substitution models, GM or homGTR.

\subsubsection{Mixture data}
\begin{figure}[]
 \begin{center}
 \includegraphics[scale=0.4]{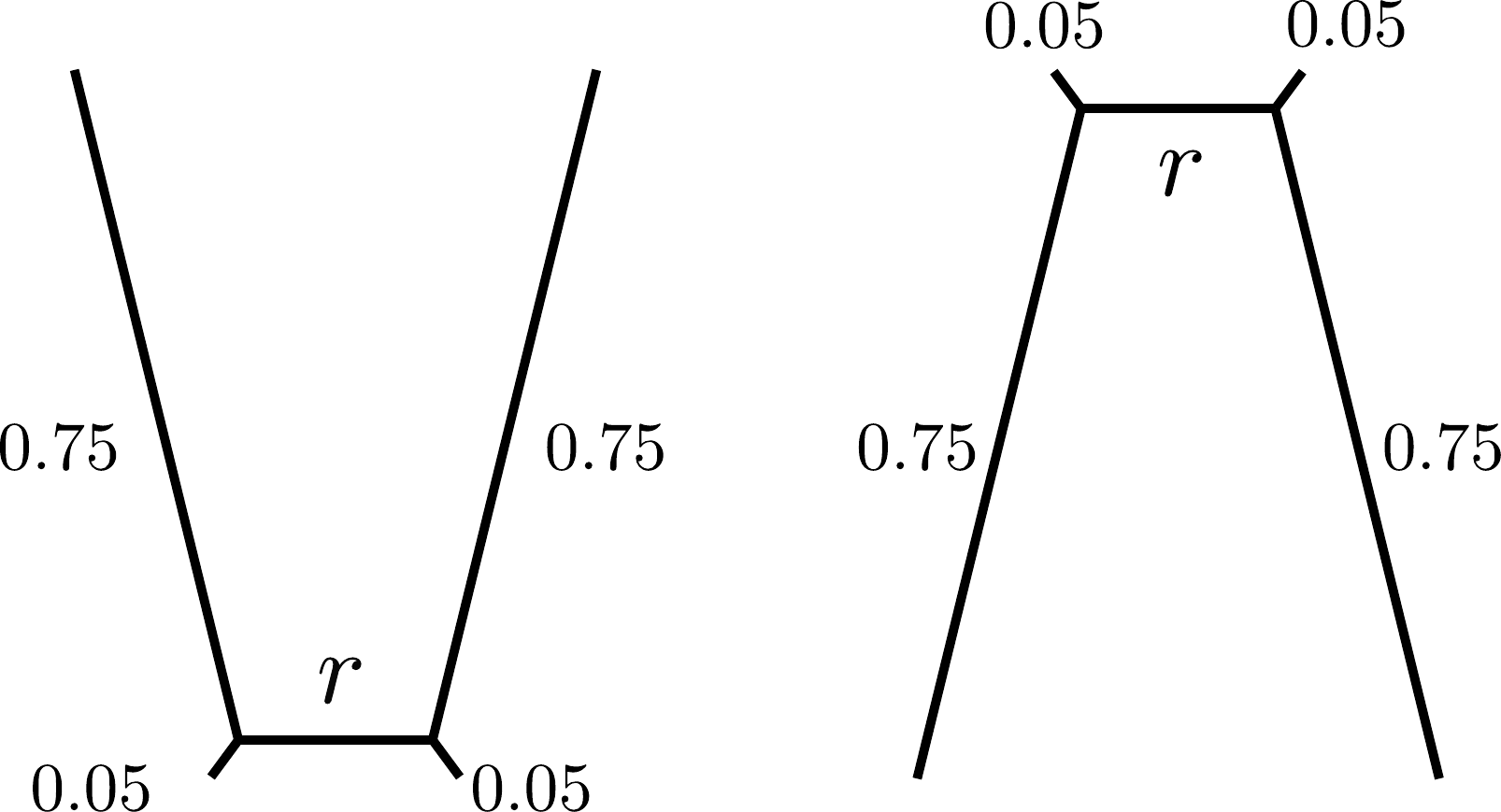}
 \end{center}
 %\vspace{3mm}
\caption{\label{kolaczkowski} \footnotesize
Mixture data taken from \cite{Kolaczkowski2004}: two categories of the same size are considered, both evolving under the GM model on the two trees depicted above with the branch lenghts indicated. The internal branch length takes the same value
in both categories and varies from 0.01 to 0.4 in steps of 0.05.
}
\end{figure}

In order to test the performance of \saq when the  assumptions underlying the method are violated, we have taken the approach by \cite{Kolaczkowski2004} and considered mixtures of distributions as follows. We take two
categories of the same sample size both evolving under the GM model on the tree of Figure
\ref{tree}.a: the first category corresponds to branch lengths $a = 0.05$, $b = 0.75$, while the second corresponds to $a = 0.75$ and $b = 0.05$ (see Figure \ref{kolaczkowski}). The internal branch length takes the same value
in both categories and varies from 0.01 to 0.4 in steps of 0.05 (see figure \ref{kolaczkowski}).

%\subsubsection{Farris tree}

%\subsubsection{Mixture data}

%

%--------Jesus: aqui va el par?graf sobre les dades reals, que v?rem dir que no inclourem en aquest article
%
%\subsubsection{Real data}
%%
%We considered the data provided by \citet{jayaswal2014} with 42~337 second codon positions of 106 orthologous genes of \textit{Saccharomyces
%cerevisiae}, \textit{S. paradoxus}, \textit{S. mikatae}, \textit{S. kudriavzevii}, \textit{S. castellii}, \textit{S. kluyveri}, \textit{S. bayanus}, and \textit{Candida albicans}.
%%
%In \citet{Rokas2003} the phylogenetic tree $T$ obtained for these species was identified with 100\% bootstrap support for the concatenated alignment of these genes.
%Although this tree is generally accepted in literature, its correct inference is known to depend on  the consideration of heterogeneity across lineages \citep{Rokas2003, Phillips2004, jayaswal2014}.  \citet{Phillips2004} obtain an alternative tree $T'$ with 100\% bootstrap using the method of minimum evolution, but identified the incorrect handling of compositional bias as responsible for this inconsistency. Moreover, according to \citet{jayaswal2014} these data are best modeled by taking into account heterogeneity across lineages plus two different rate categories $2\Gamma$ and invariable sites $I$.

\newpage
\section{Results}
% !TEX root = ms.tex

%\private{Presentem els resultats obtinguts en aplicar el diferents metodes, organitzat pel tipus de dades on ho apliquem. Tambe descrivim el temps d'execucio.  La web o la referencia d'on trobar el codi dels nous metodes. }

\subsection{Tree space}

\begin{figure}[H]
	\begin{center}
		\includegraphics[scale=0.5]{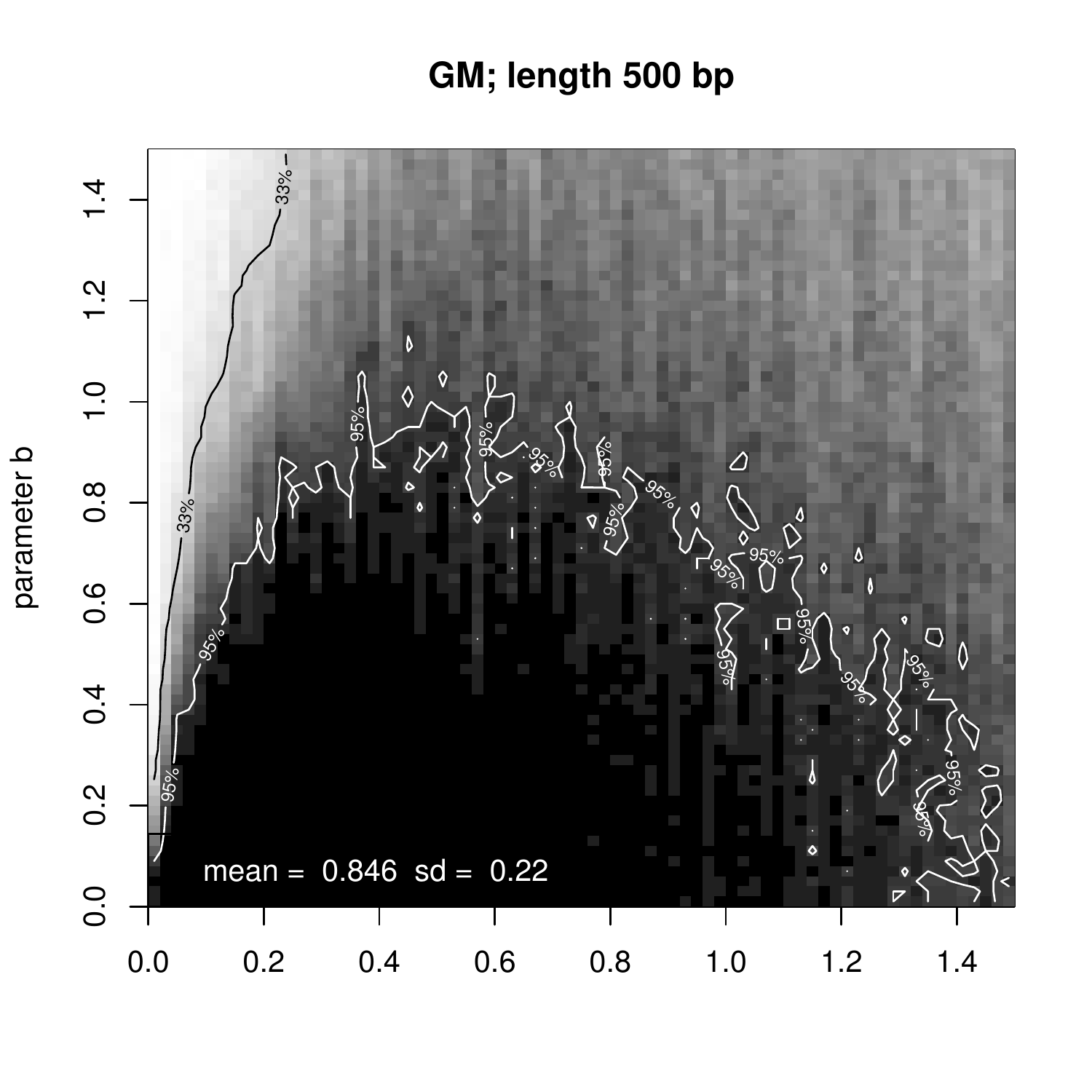}
		\includegraphics[scale=0.5]{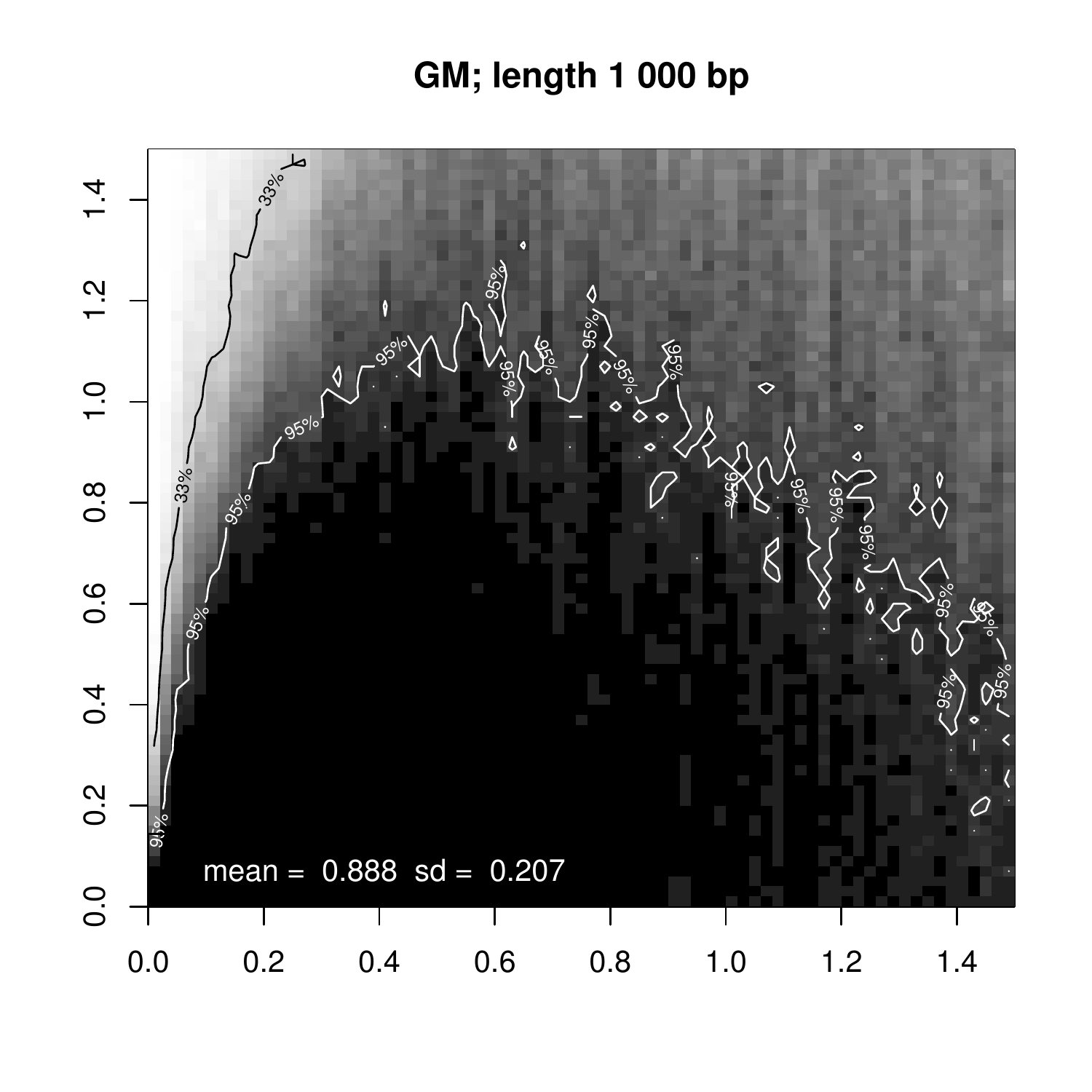}\\
		\includegraphics[scale=0.5]{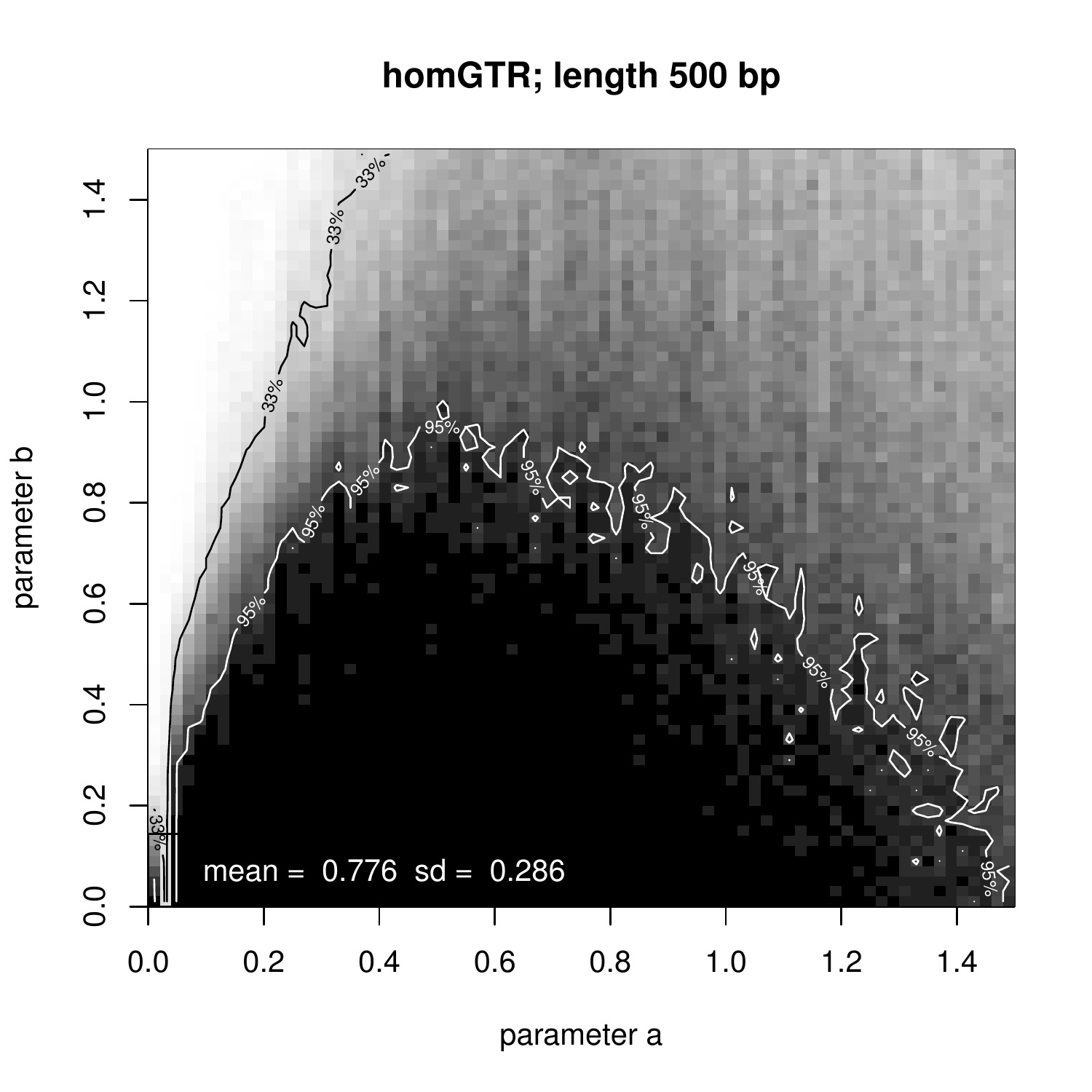}
		\includegraphics[scale=0.5]{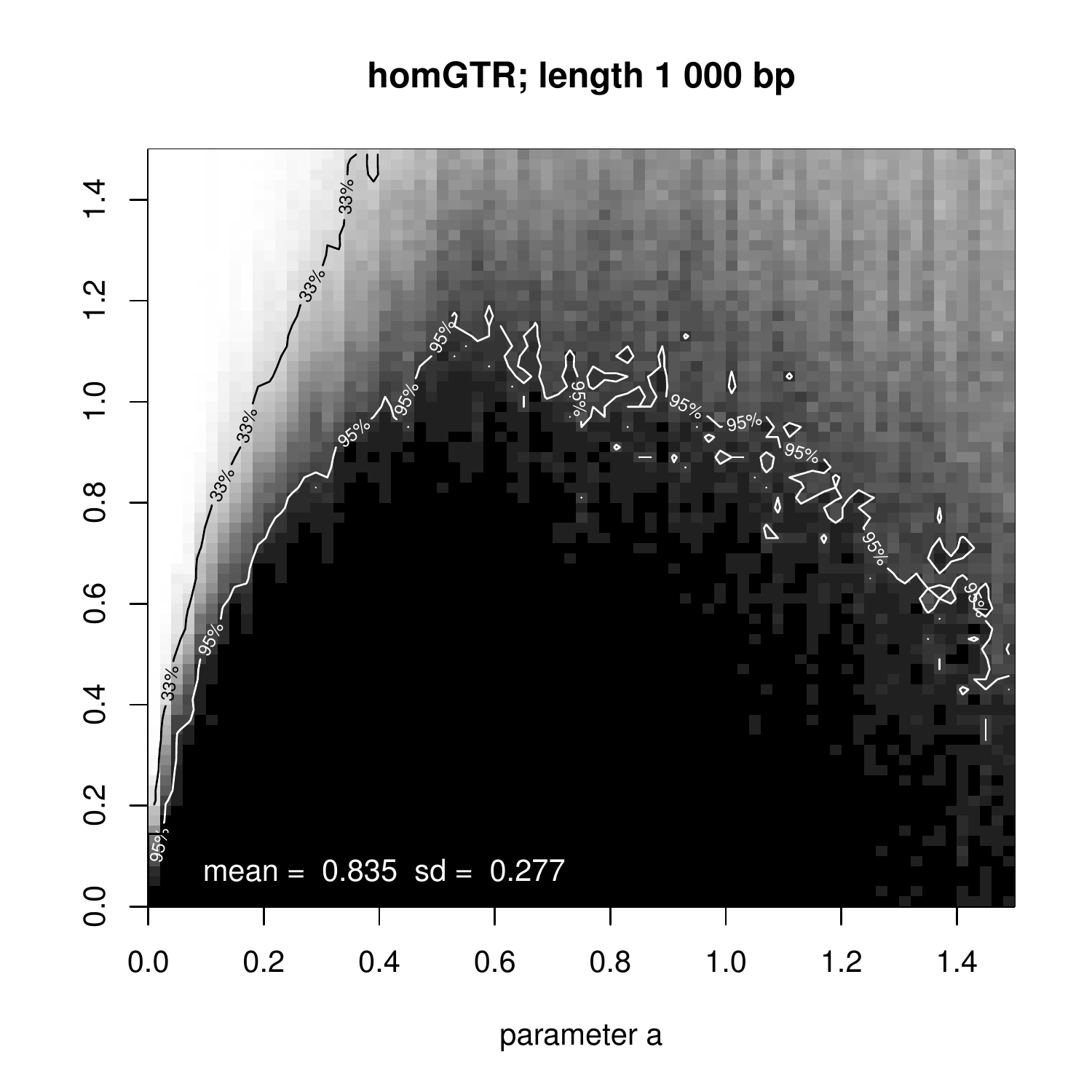}
	\end{center}
	%\vspace{3mm}
	\caption{\label{treespaces} \footnotesize
	Performance of \saq in the tree space of Figure \ref{tree}.b on alignments of length  500 bp (left) and 1 000 bp (right).  Black is used to represent 100\% of successful quartet reconstruction,
white to represent 0\%, and different tones of gray the intermediate frequencies. The 95\% contour line
is drawn in white, whereas the 33\% contour line is drawn in black. Top: data generated under the GM model; Bottom: data generated under a homogenous GTR model.
}
\end{figure}

The performance of \saq on the tree space data for alignments of 500 and 1 000 bp.  simulated under GM or homGTR data is presented in figure \ref{treespaces}. The success of \saq in recovering the correct quartet is represented by different tones of gray, where black corresponds to a 100\% success and white corresponds to a 0\% success. Gray tones correspond to regions of intermediate probability, and the 95 \% and 33 \% isoclines are represented with a white line.

We observe that these figures exhibit a consistent performance (according to the results by \citet{huelsenbeck1995} and \cite{casfer2016}), with a decreasing performance at the Felsenstein zone but a high performance at the other zones, and with an increase of success for larger samples.
In summary, the average success of \saq applied to alignments of length 10 000 bp. is 96.8 \%  when applied to GM data, and 94.5 \% when applied to homGTR data.
%(standard deviation of 0.14) - (standard deviation of 0.21)

%The average success of \saq obtained in the data under the general Markov model (GM) or the time-reversible model homogeneous across lineages and sites (homGTR) for different lengths (see figures ???) is shown in the following table:

 %\begin{center}
 % \begin{tabular}{c|ccc|ccc}
 %\hline
%model & GM &  &  & homGTR & & \\ \hline
%b.p. & 500 & 1000 & 10000 & 500 & 1000 & 10000 \\ \hline \hline
%  & 0.840 (0.21) & 0.877 (0.20) & 0.952 (??) &  0.794 (0.24) & 0.842 (0.22) &  0.937 (??) \\\hline
%   \end{tabular}
%   \end{center}
%In parentheses we show the standard deviation of the set of percentages of success of each method in each %tree space. This figures can be easily confronted to
% those obtained by \cite{casfer2016} for \eri and other methods, and they exhibit

% TABLE

%\subsection{Comparison to other methods}\blue{cal posar-hi titol? Despres tambe comparem mixtures amb altres metodes} \tp{realment, crec que no cal}

In Table \ref{tab:mean_sd} we summarize the overall performance of \saq on the tree space in comparison with the methods studied \citep{casfer2016}. In particular, we compare it with a maximum likelihood approach: ML(homGMc) estimates the most general homogeneous across lineages continuous-time model and ML(homGTR) estimates a homogeneous across lineages GTR model (both methods estimate the rate matrix and the distribution at the root). We also compare it to the neighbor-joining method (NJ) under the paralinear distance, as was used in the quoted paper. We also provide the comparison to the algebraic method \eri, which is based on a GM model and makes only use of condition (b) explained in the Methods section. The plots of the performance of these methods on the tree space described above can be found in \cite{casfer2016}; here we summarize the results in Table \ref{tab:mean_sd}.

\begin{table}[H]
\begin{center}
\textbf{Average success of different quartet methods on the tree space of Figure \ref{tree}b.}
\vspace*{5mm}

%  \begin{tabular}{cc|cccc}
%  \hline
%   simulations & base pairs & \saq & %\svd &
%   \eri & \nj & \ml \\ \hline \hline
%   GM & 500  & 0.846 (0.22) & %0.800 (0.23) &
%    0.724 (0.20)  &  0.725 (0.21) & 0.721 (0.19)\\
%   & 1~000  & 0.888 (0.21) & %0.856 (0.21) &
%   0.803 (0.17) & 0.797 (0.18) & 0.736 (0.17)\\
%   %& 10 000  & 0.968 (0.14) & %0.958 (0.13) &
%   %\textbf{0.971} (0.04) & 0.943 (0.09) &  0.754 (0.17) \\
%   \hline
%   homGTR  & 500 & 0.784 (0.29) & %0.732 (0.21) &
%   0.748 (0.22) & 0.729 (0.23) & 0.880 (0.11) \\
%   & 1~000 & 0.835 (0.28) & %0.796 (0.30) &
%   0.843 (0.19) & 0.805 (0.20) & 0.934 (0.06) \\
%   %& 10 000 & 0.945 (0.21) & %0.940 (0.22) &
%   %\textbf{0.992} (0.04) & 0.945 (0.10) & 0.980 (0.02) \\
%   \hline
%    \end{tabular}

 \begin{tabular}{cc|cccc}
 \hline
  simulations & base pairs & \saq & %\svd &
  \eri & \nj & \ml \\ \hline \hline
  GM & 500  & {84.6}  & %0.800 (0.23) &
   72.4   &  72.5  & 72.1\\
  & 1~000  & {88.8}  & %0.856 (0.21) &
  80.3  & 79.7  & 73.6 \\
  %& 10 000  & 0.968 (0.14) & %0.958 (0.13) &
  %\textbf{0.971} (0.04) & 0.943 (0.09) &  0.754 (0.17) \\
  \hline
  homGTR  & 500 & 78.4 & %0.732 (0.21) &
  74.8  & 72.9  & {88.0}  \\
  & 1~000 & 83.5  & %0.796 (0.30) &
  84.3 & 80.5  & {93.4} \\
  %& 10 000 & 0.945 (0.21) & %0.940 (0.22) &
  %\textbf{0.992} (0.04) & 0.945 (0.10) & 0.980 (0.02) \\
  \hline
   \end{tabular}

% \begin{tabular}{cc|cccc}
%	\hline
%	simulations & base pairs & \saq & %\svd &
%	\eri & \nj & \ml \\ \hline \hline
%	GM & 500  & \textbf{84.6} (0.22) & %0.800 (0.23) &
%	72.4 (0.20)  &  72.5 (0.21) & 72.1 (0.19)\\
%	& 1~000  & \textbf{88.8} (0.21) & %0.856 (0.21) &
%	80.3 (0.17) & 79.7 (0.18) & 73.6 (0.17)\\
%	%& 10 000  & 0.968 (0.14) & %0.958 (0.13) &
%	%\textbf{0.971} (0.04) & 0.943 (0.09) &  0.754 (0.17) \\
%	\hline
%	homGTR  & 500 & 78.4 (0.29) & %0.732 (0.21) &
%	74.8 (0.22) & 72.9 (0.23) & \textbf{88.0} (0.11) \\
%	& 1~000 & 83.5 (0.28) & %0.796 (0.30) &
%	84.3 (0.19) & 80.5 (0.20) & \textbf{93.4} (0.06) \\
%	%& 10 000 & 0.945 (0.21) & %0.940 (0.22) &
%	%\textbf{0.992} (0.04) & 0.945 (0.10) & 0.980 (0.02) \\
%	\hline
%\end{tabular}

   \end{center}

\caption{\label{tab:mean_sd} \footnotesize Average success of \saq corresponding to the tree space of Figure \ref{tree} b, compared with the results of the performance of \eri, neighbor-joining (\nj) and  maximum likelihood (\ml) taken from \citep{casfer2016}. %In parentheses we show the standard deviation of the set of percentages of success of each method in each tree space. In each row, the highest success is indicated in bold font. 
\mll(homGMc) estimates a homogeneous continuous GM model (that is, it estimates an unrestricted rate matrix for the whole tree and a distribution at the root) and is applied when data is generated under a GM model, while \mll(homGTR) is applied when data are generated under homGTR.}
\end{table}

\newpage
\subsection{Random branch lengths}

\begin{figure}
 \begin{center}
\includegraphics[scale=0.15]{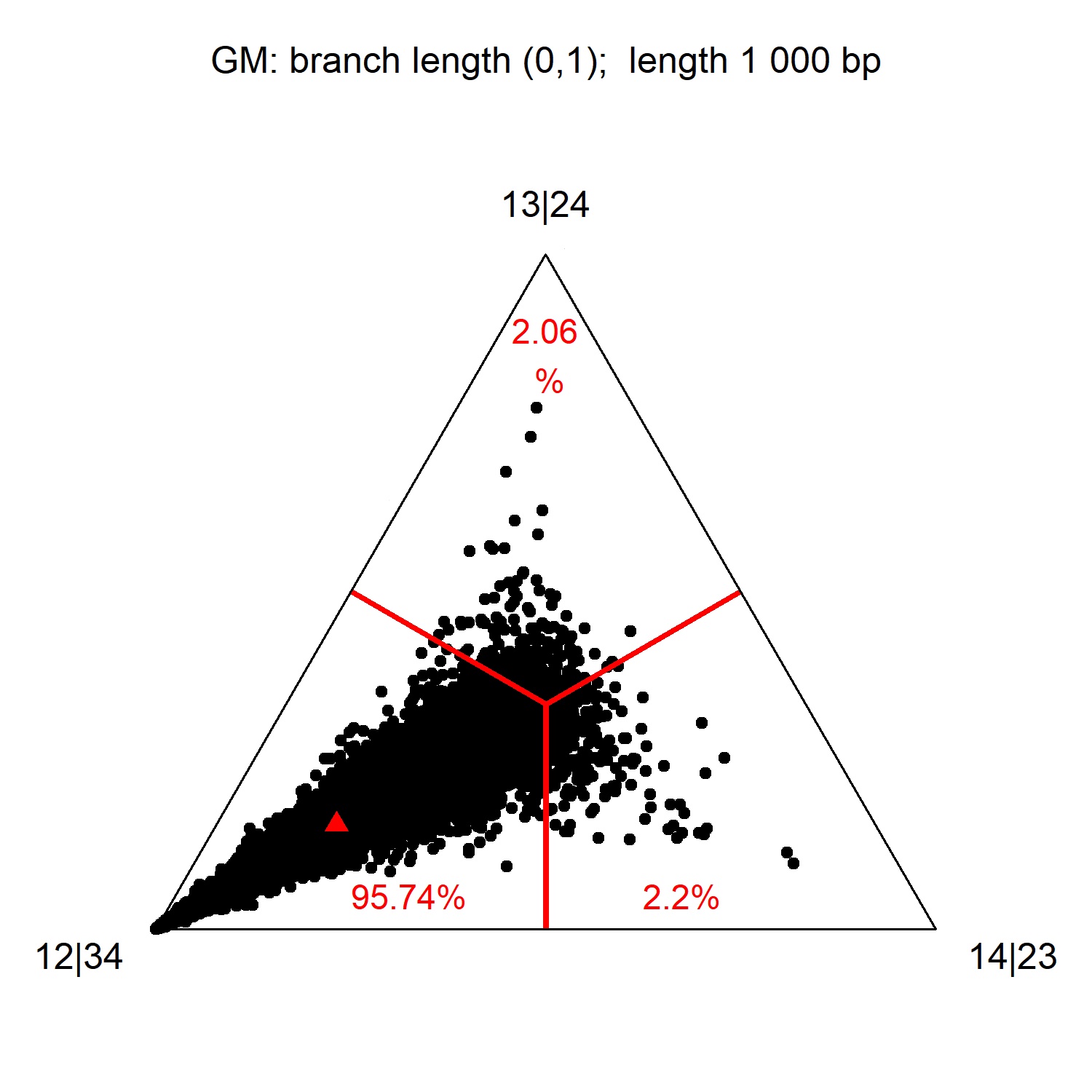}
\includegraphics[scale=0.15]{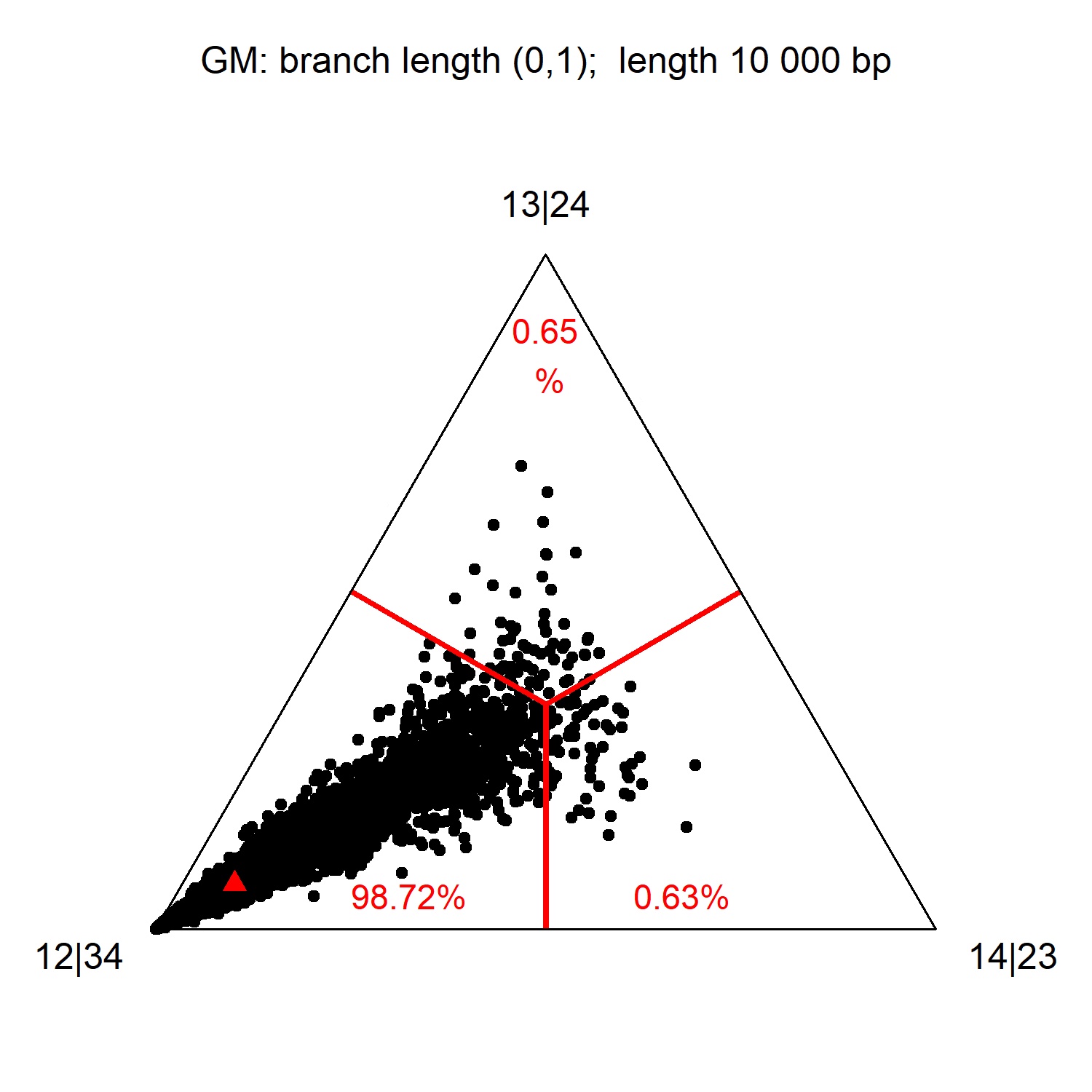} \\
\includegraphics[scale=0.15]{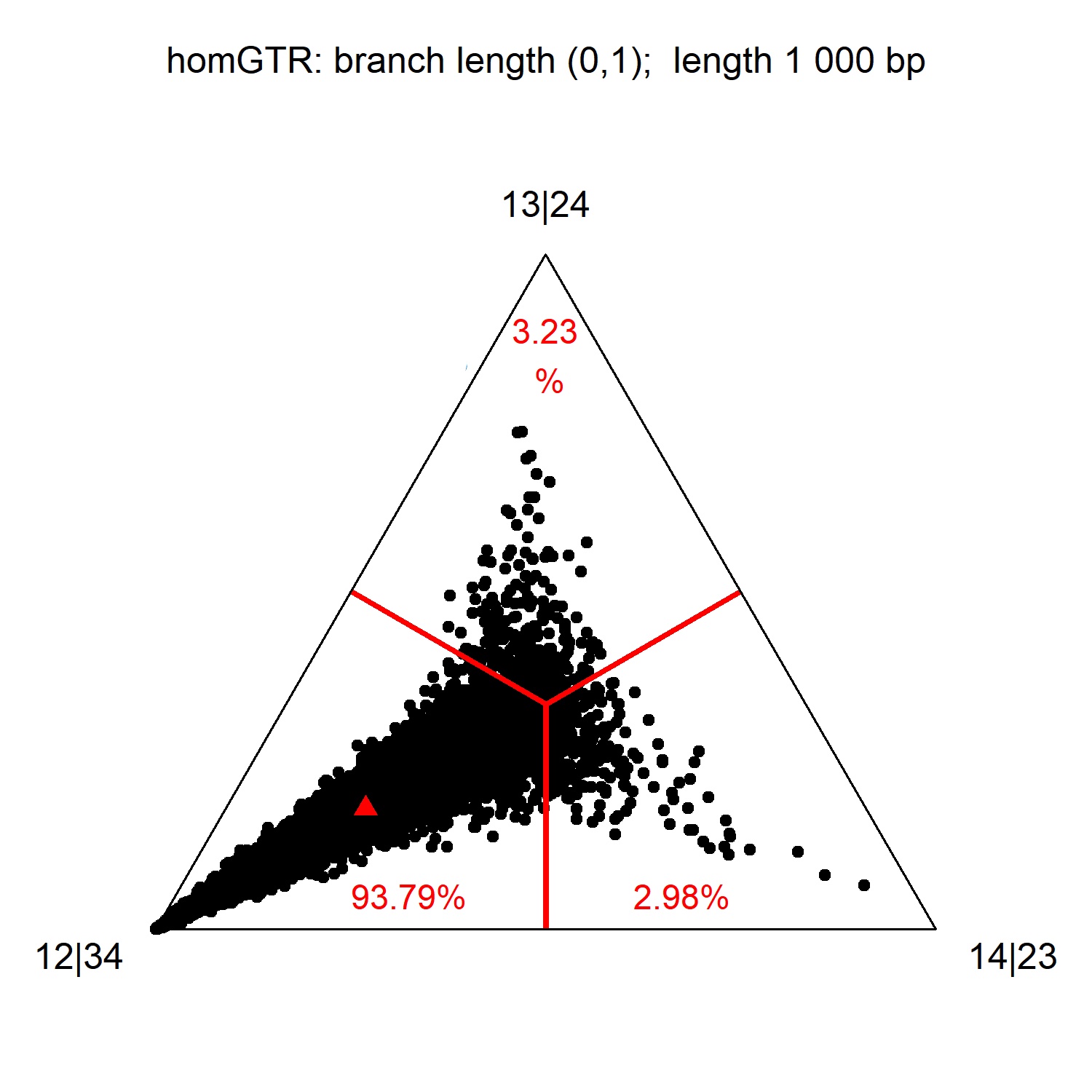}
\includegraphics[scale=0.15]{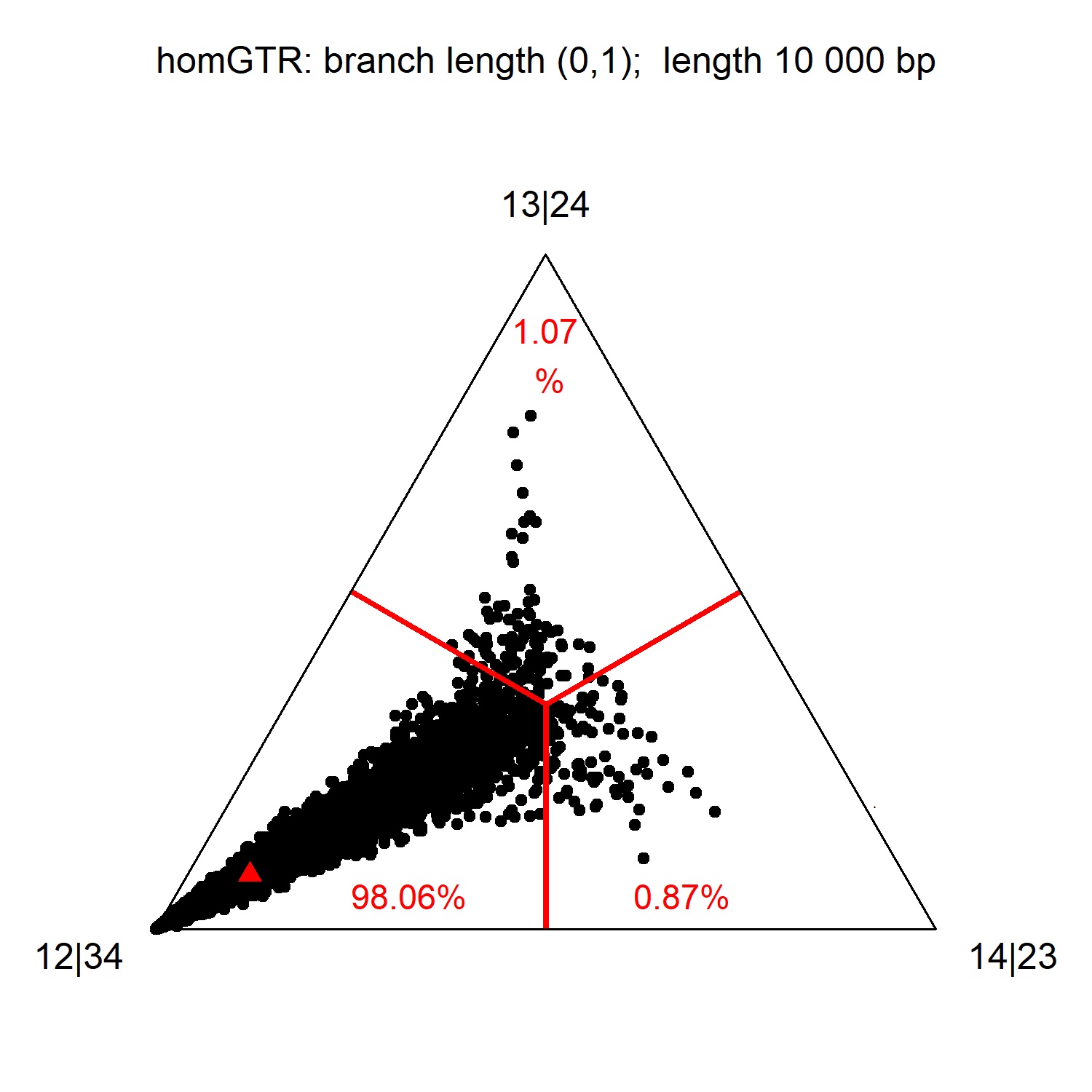}
  \end{center}
 %\vspace{3mm}
\caption{\label{ternary_01} \footnotesize
%\tp{PENDENT: editar figures}
Ternary plots corresponding to the weights of \saq applied to 10000 alignments
generated under the $12|34$ tree with random branch lengths uniformly distributed between
0 and 1. On each triangle the bottom-left vertex represents the underlying tree
$12|34$, the bottom-right vertex is the tree $13|24$ and the top vertex is $14|23$.
Top: correspond to data generated under GMM; bottom: data
generated under GTR. Left : 1000 bp; Right: 10000 bp.}
\end{figure}

As the weights of \saq are normalized so that
the three quartet values sum to one,  it is suitable for input of quartet-based methods and also allows plotting the scores in a ternary
plot (also called a simplex plot), see \citep{strimmer1997}. To visualize how the output of \saq is distributed, we show ternary plots corresponding to the alignments generated on the tree $12|34$ under the GM and homGTR models, with lengths 1~000 bp. and 10~000 bp. and random branch lengths uniformly distributed in (0,1) (Figure \ref{ternary_01}). The ternary plots of the performance of \saq when applied to the same setting with branch length uniformly distributed in (0,3) is shown in Figure \ref{ternary_03} of the Appendix.

% ternary plot GMM
%
% We observe that \saq scores correctly distributed, with the majority points close to the left corner (which represents the correct topology) for both GM and homGTR data.
%
We observe that \saq weights are equally distributed for the two wrong topologies and the vast majority of points lie close to the left corner (which represents the correct quartet) for both GM and homGTR data.
We also observe a strong difference in the performance of \saq when applied to branch lengths in (0,1) or in (0,3), being notable higher in the former. 
% \blue{NO afegim comparacio a ML? Al Syst bio no vam posar els resultats} \texttt{ML}(homGM) applied to the same data under GM obtained a success of 50.14 (length 1 000 bp.) and 51.36 (length 10 000 bp.), while the success of \texttt{ML}(homGTR) for GTR data was remarkably higher: 78.19 (length 1 000 bp.) and 92.34 (length 10 000 bp.). 
%
The average success of \saq for these random branches systems is shown in Table \ref{tab:random_branches}.

\vspace{3mm}
\begin{table}[H]
\textbf{Average success of \saq applied to data generated on $12|34$ with random branch lengths}
%\vspace*{5mm}
\begin{center}
%  \begin{tabular}{c|c|c|}
%  GM & 1000 & 10000 \\ \hline \hline
%  (0,1) & 95.74 & 98.72 \\
%  (0,3) & 69.74 & 82.69 \\ \hline
%    \end{tabular}
%    \qquad \qquad
%    \begin{tabular}{c|c|c|}
%   homGTR & 1000 & 10000 \\ \hline \hline
%  (0,1) & 93.72 & 98.06 \\
%  (0,3) & 67.87 & 84.51 \\
%  \hline
%    \end{tabular}
\begin{tabular}{ccccc}
\multicolumn{1}{l}{}       & \multicolumn{2}{c}{GM}                      & \multicolumn{2}{c}{homGTR}                  \\ \cline{2-5}
\multicolumn{1}{c|}{}      & 1 000 bp. & \multicolumn{1}{c|}{10 000 bp.} & 1 000 bp. & \multicolumn{1}{c|}{10 000 bp.} \\ \hline
\multicolumn{1}{c|}{(0,1)} & 95.74     & \multicolumn{1}{c|}{98.72}      & 93.72     & \multicolumn{1}{c|}{98.06}      \\
\multicolumn{1}{c|}{(0,3)} & 69.74     & \multicolumn{1}{c|}{82.69}      & 67.87     & \multicolumn{1}{c|}{84.51}      \\ \hline
\end{tabular}
\end{center}

\caption{\label{tab:random_branches} \footnotesize Average success of \saq on alignments of lengths 1000 and 10000 bp. generated on the tree $12|34$ under the GM and homGTR models with random branch lenghts uniformely distributed in (0,1) (first row) and (0,3) (second row). }
\end{table}

\subsection{Mixture data}

\begin{figure}[H]
	\begin{center}
		\includegraphics[scale=0.13]{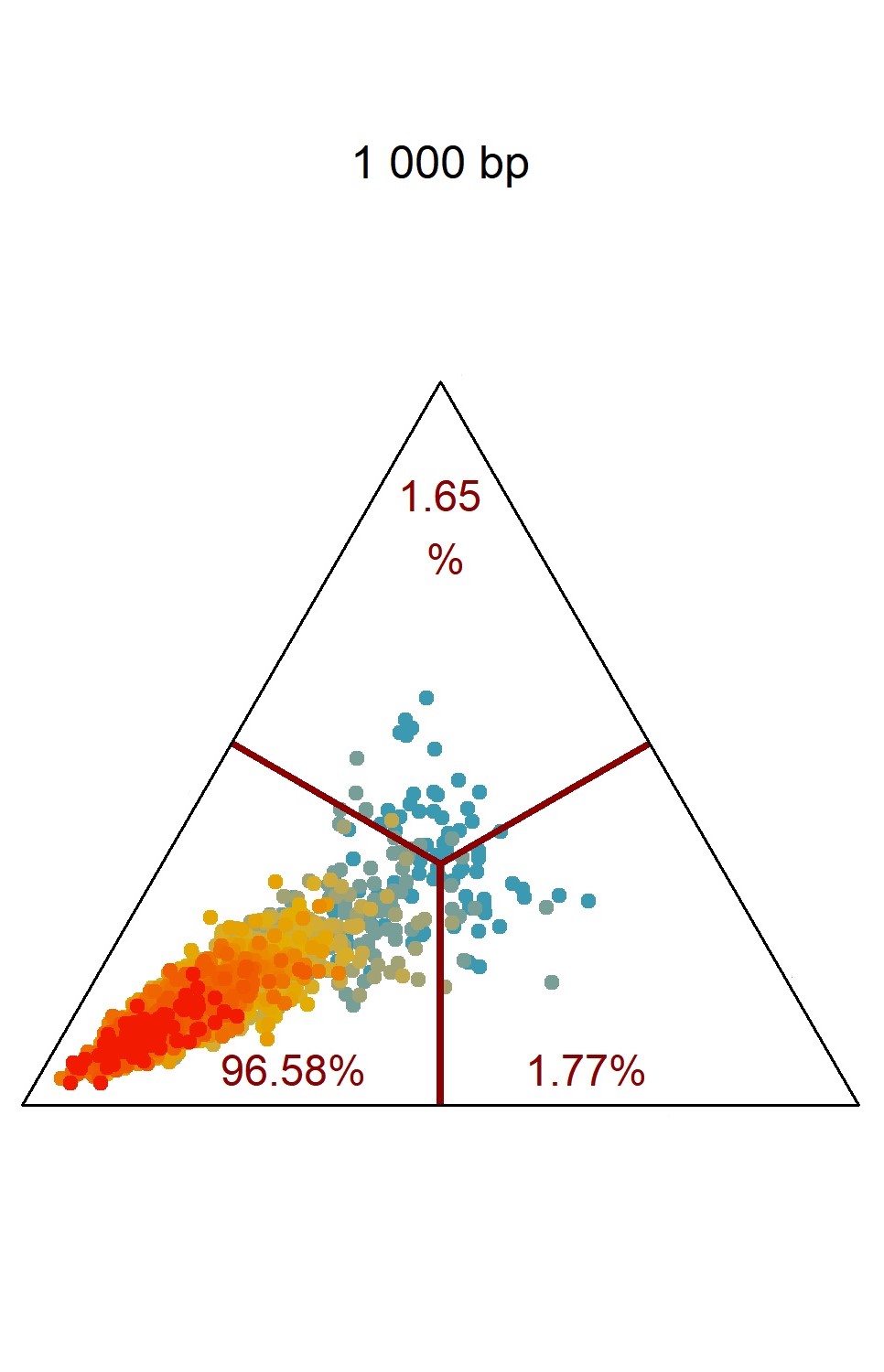}
		\includegraphics[scale=0.13]{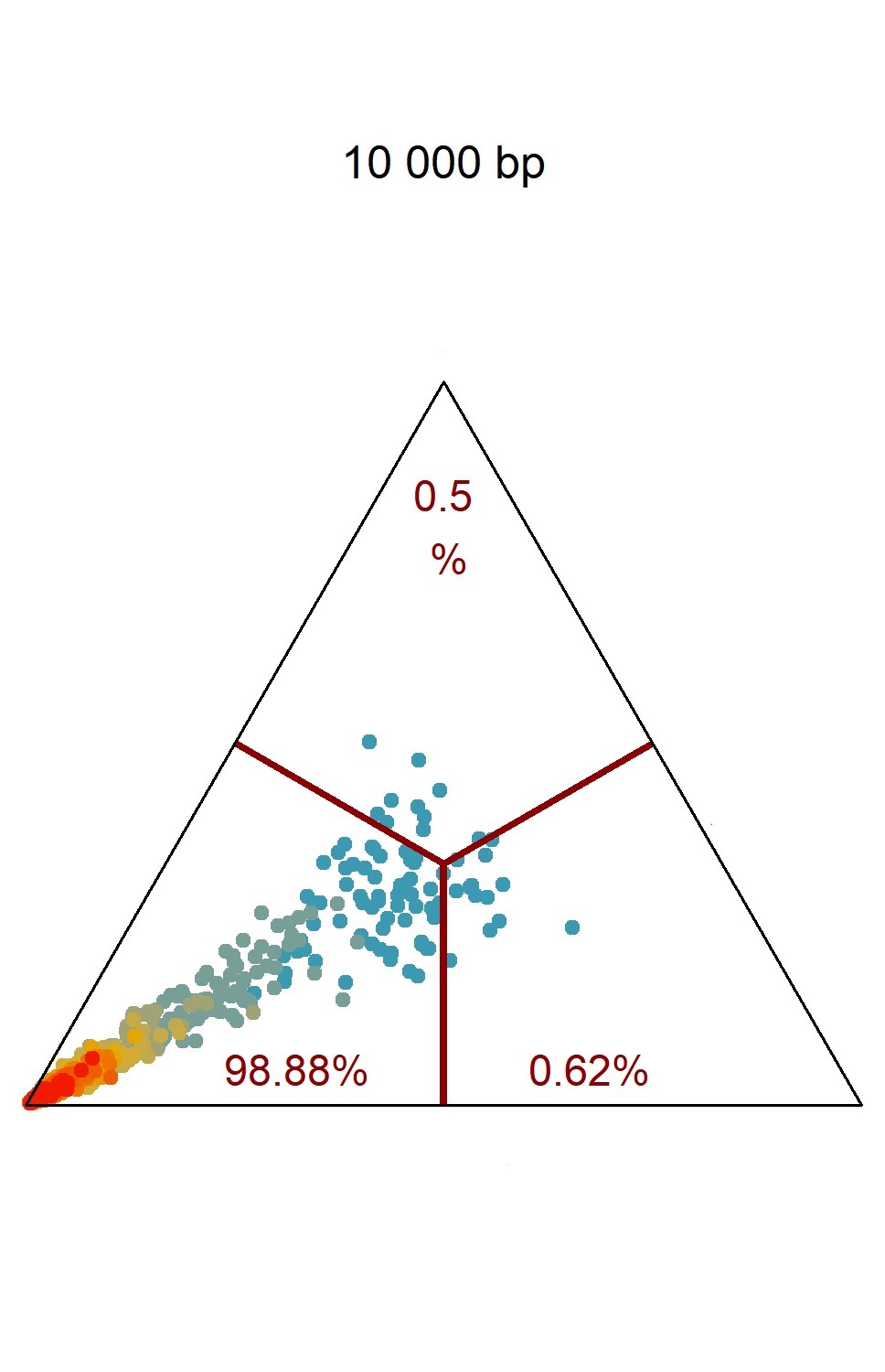}
		\includegraphics[scale=0.13]{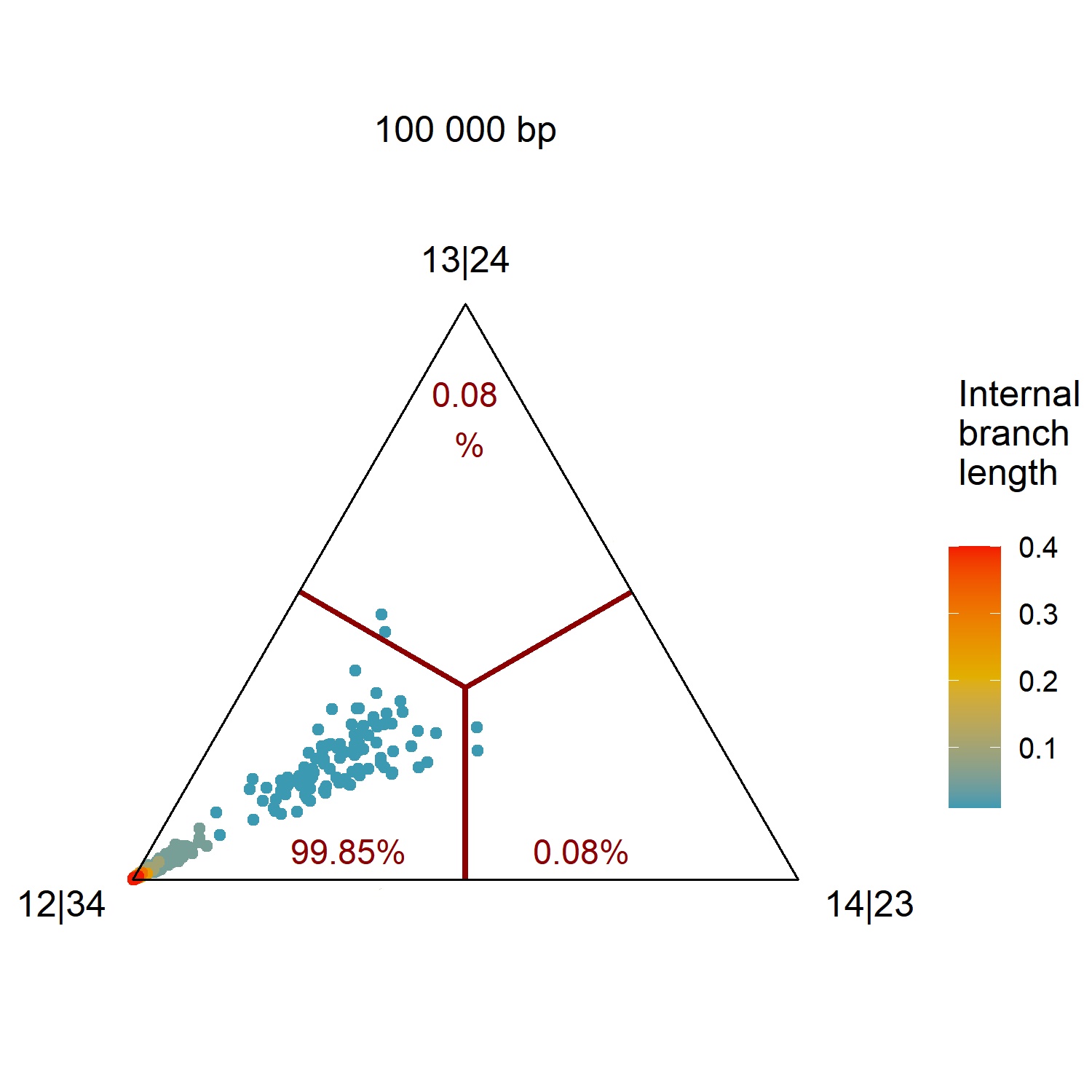}\\
		\includegraphics[scale=0.65]{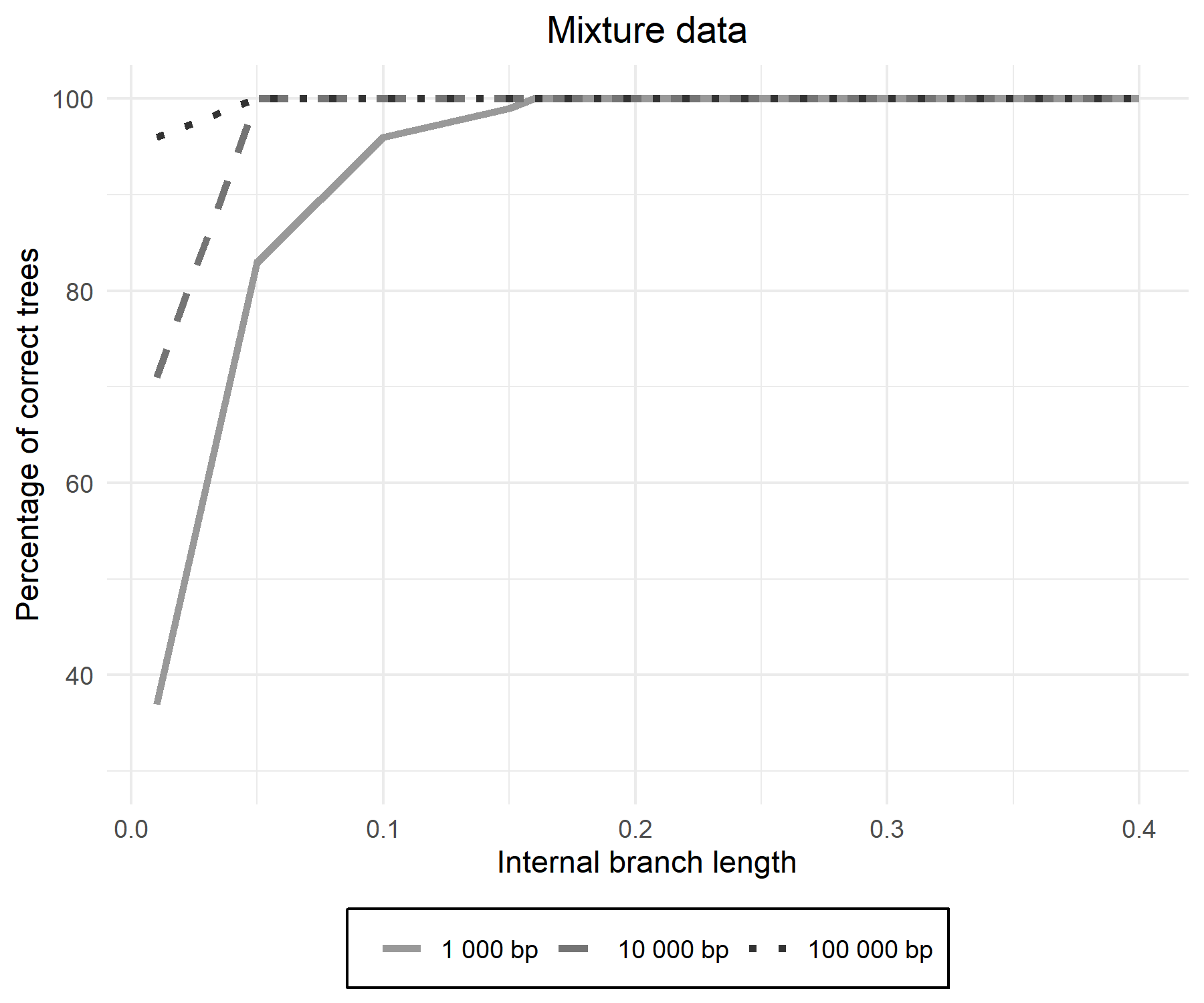}
		
	\end{center}
	%\vspace{3mm}
	\caption{\label{ternary_mixtures} \footnotesize The three plots on the top show the ternary diagrams corresponding to the weights of \saq applied to heterogeneous alignments of lengths 1 000, 10 000 and 100 000 bp. generated by the trees of 
		% 	with two categories according to the test designed in (Kolaczkowski
		% and Thornton 2004)  corresponding to the $12|34$ trees of 
		figure \ref{kolaczkowski}. On each triangle the bottom-left vertex represents the underlying true topology
		$12|34$, the bottom-right vertex is topology $13|24$ and the top vertex is $14|23$. 
		Each triangle is divided into three regions according to which tree is selected by the method. The figures in red represent  the percentage of alignments of the corresponding region according to \saq. 
		The gradient in color of the dots describe the variation of the branch length $r$ of the interior edge, from 0.01 to 0.4 in steps of 0.05, as indicated by the bar on the right. 
		For the same data, the plot on the bottom represents the percentage of correctly reconstructed trees by \saq as a function of the internal branch lenght.  Data
		generated under the GM model with 2 categories, varying the internal branch length, and recovering with \saq. }
\end{figure}

The performance of the method \saq under the mixed data described in the Methods section (see also figure \ref{kolaczkowski}) is shown in figure \ref{ternary_mixtures}. The ternary diagrams show a high accuracy in determining the correct quartet, even when the length of the alignments is 1 000 bp. In the same figure, we present the performance of \saq in terms of the branch length of the interior edge, following the study suggested by \cite{Kolaczkowski2004}, for lengths 1 000, 10 000 and 100 000 bp. This plot is to be compared with the analogous plot in \cite{casfer2016} and we summarize the comparision to other methods in Table \ref{tablemixtures}.  As it is apparent from the results in the table, \saq outperforms all the other methods (even \eri (2)) despite mixture data violates the assumptions of this method.

\begin{table}[H]
\textbf{Performance of different methods applied to mixture data}
%\vspace*{5mm}
\begin{center}
\begin{tabular}{c||ccccc}
      %& \multicolumn{5}{c}{1000}     \\
internal  branch length & 0.01 & 0.05 & 0.1 & 0.2 & 0.3  \\ \hline \hline
\texttt{SAQ}                                                               & 37   & 83   & 96  & 100 & 100 \\
%\texttt{Erik+2}                                                            & 5    & 14   & 25  & 76  & 99  & 2      & 2    & 10  & 68  & 98  \\
\texttt{Erik+2 (2)}  & 12   & 35   & 60  & 86  & 96  \\
\texttt{MP}                                                                & 0    & 2    & 19  & 76  & 99  \\
\texttt{ML}(GTR+2 $\Gamma$) & 0 & 4 & 14 & 77 & 95
\end{tabular}
\end{center}

\caption{\label{tablemixtures} \footnotesize Percentage of correctly reconstructed topologies by different methods on alignments of length 1 000 bp. for data
generated under the GM model with 2 categories according to the test designed described in the section data and varying the internal branch length, and recovering with \saq, \eri with 2 partitions, Maximum Parsimony and \texttt{ML}(GTR+2 $\Gamma$) estimating time-reversible model with 2 discrete-gamma categories. 
For all internal branch lenghts, ML had to estimate all parameters. 
}
\end{table}

%\subsection{Heterogeneity across sites }

%\subsection{Performance on real data}

\subsection{Execution time}
The computations were performed on a machine with 6 Dual Core Intel(R) Xeon(R) E5-2430 Processor (2.20 GHz) equipped with 25 GB RAM running Debian GNU/Linux 8. We have used the g++ (Debian 4.9.2-10+deb8u2) 4.9.2 compiler and the C++ library for linear algebra \& scientific computing \textit{Armadillo} version 3.2.3 (Creamfields).

The time needed to apply \saq to $100$ alignments of length 10~000 bp is $9$ seconds.

\section{Discussion and conclusions}
% !TEX root = ms.tex

In this paper, we have presented a new quartet reconstruction method that we call \saq. It is a robust and accurate method, which is essentially based on the results obtained by \cite{AllmanSemialg} and makes profit of the stochastic information available in the data to infer the topology of the phylogenetic tree. As far as we are aware, this is the first method that combines both the algebraic and the semialgebraic nature of the underlying substitution models.

The study and the simulations carried out in this paper show that \saq is a robust phylogenetic reconstrucion method, specially when dealing with data generated under the GM model. 
We have even proven that the performance of the method is high for data that violate the hypothesis of the model (mixture data).
%the performance of the method. The performance under a homogeneous GTR model is not that good, but pretty remarkable, too.
In connection with the performance on mixed data, we would like to point out that although \saq is not specifically designed to deal with mixtures of distributions in general, it can be easily adapted to deal with invariable sites. One only needs to switch the relative frequency vector $p$ by a corrected vector that takes into account the estimated amount of invariable sites (e.g. as in \citet{jayaswal2007} or \citet{steel2000}). An implementation of this feature in \saq is on the way.

The results described in the preceeding section show that the method performs better when applied to data generated under the GM model than under a homogeneous GTR model (for which the results are satisfactory and similar to the method \eri). 
% One possible explanation for this phenomena is that on GTR data generated on the tree $12|34$ with a transition matrix at the interior edge close to the identity (so that the method is pushed to the limit), the flattening matrices $Flat_{12|34}(\tilde{p}_i)$ are close to be symmetric and positive definite (at least when the stationary distribution is uniform). In this situation, the method practically relies on the distance to rank 4 matrices and as a consequence, the results are similar to those of \eri.
One possible explanation for this phenomena is that on GTR data, when the transition matrix at the interior edge is close to the identity (so that the method is pushed to the limit), the flattening matrices $F_{12|34}$ needed to compute the weights of \saq are close to be symmetric and positive definite (at least when the stationary distribution is uniform). In this situation, the semialgebraic information of the data become almost irrelevant and the method practically relies on the distance to 4-rank  matrices. As a consequence, the results are similar to the results of other methods based on this distance, as \eri.

% * The results for lenth 10000 are not as good as we would like...\blue{Posar-ho? Potse no cal}
%Overall, the method will perform slightly worse than for more general data. CHECK!!!

% The idea of the \saq method To this aim, it applies a number of transformations to the distributon vector obtained from a 4-species alignment, producing other vectors consistent with the same tree. The stochastic nature of these transformed vectors is then used to reveal the right topology.

\saq is developed for quartets and assigns a normalized weight between 0 and 1 to each possible tree: this weight is larger to the tree with the higher confidence. We have checked that these normalized weigths are unbiased (in the sense that there is no trend towards any of the incorrect trees) and are statistically consistent. % (they clearly point to the correct tree when the amount of data increases).
Although \saq is only developed for quartets, its weights can be used as input of quartet-based methods such as ``weight optimization'' by \cite{Ranwez2001}, ``quartet-puzzling'' by \cite{strimmer1996} or the method by \cite{willson}. We plan to test the weights of \saq as input of quartet-based methods in a forthcoming work.

% Finally, let us note that combining algebraic and semi-algebraic conditions in a single score has been a tedious task. There could be many different ways of combining all the restrictions satisfied by the theoretical distributions. For example, in the definition of the score $s_T^i$, we could have considered not projecting into symmetric positive matrices for the flattening $F_{12|34}$ as this seems as adding unnecessary complexity to the algorithm. However, we found that by considering the projection, all $\delta_4$ are computed on the same space of positive definite matrices and improves the results of the algorithm. Analogously, we have tested many other options that are not mentioned in this final report.

Finally, let us note that combining algebraic and semi-algebraic conditions in a single score has been a tedious task. There are many different ways of combining all the restrictions satisfied by the theoretical distributions. For example, in the definition of the score $s_T^i$, we could have considered not projecting into symmetric positive matrices for the flattening $F_{12|34}$ as this seems as adding unnecessary complexity to the algorithm. However, we found that by considering the projection, all distances to 4-rank matrices are computed on the same space of positive definite matrices and improves the results of the algorithm. Analogously, we have tested many other options that are not mentioned in this final report.
In this sense,  \saq is the result of a number of decisions based on  exhaustive simulation studies, with the scope of obtaining a method that takes into account all the information at hand, but at the same time,  keeping the method computationally feasible and as simple as possible.
We are aware that a detailed statistical analysis would be convenient in order to justify the good performance observed in the simulations.

%\tp{observacions: when the model is GTR and $Flat_12|34(R)$ is simeq and def $+$ (veure notes Marta; potser cal dist. uniforme). If $M_5\cong Id$, then $Flat_{12|34}(\tilde{p}_i)$ are close to be symmetric and def +, so $\delta_4()$ takes in account only the rank. Maybe this is the reason why the method is not so good for GTR data...}

%\section{Acknowledgments}

\section{Finantial support}
The authors were partially supported by Spanish government Secretar\'{\i}a de Estado de Investigaci\'{o}n, Desarrollo e Innovaci\'{o}n [MTM2015-69135-P (MINECO/FEDER)] and [PID2019-103849GB-I00 (MINECO)]; Generalitat de Catalunya [2014 SGR-634]. M. Garrote-L\'{o}pez was also funded by Spanish government, Ministerio de Econom\'{\i}a y Competitividad research project Maria de Maeztu [MDM-2014-0445].

\section{Author's contributions}
All authors contributed equally.

\bibliographystyle{apalike}
%\bibliography{biblio}

\newpage
\section{Appendix}
% !TEX root = CasanellasFernandezGarrote.tex

%Descriure les transformacions dels tensors (dibuix, però no detalls tècnics)
%
%El rang 4 per als flattenings de la topologia correcte -> Allman - Rhodes; aquesta és la idea darrera ErikSVD i Erik+2
%
%ART matrius simètriques def +; fem servir la estocasticitat
%

\subsection{Insight on the theoretical basis of the method}

\subsubsection{$T$ leaf-transformations}
\begin{figure}[h]
 \begin{center}
 \includegraphics[scale=0.28]{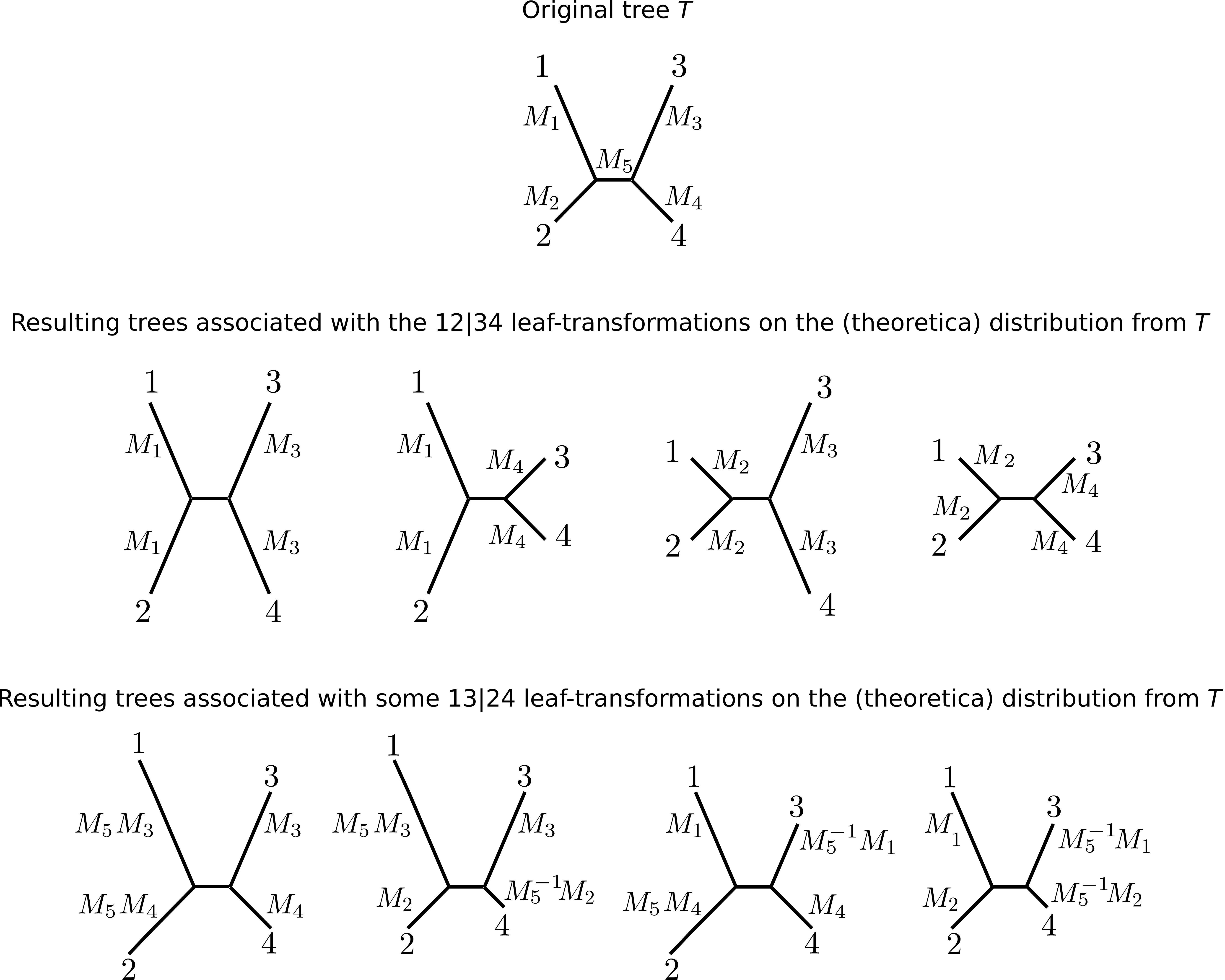}
 \end{center}
 %\vspace{3mm}
\caption{\label{transformations} \footnotesize
%Effect of the $12|34$ leaf transformations on the tree of the Figure \ref{tree} a).
For the phylogenetic tree $T=12|34$ (on the top of the figure), we show the effect on the parameters of applying the $12|34$ leaf-transformations (middle) and some $13|24$ leaf-transformations (bottom) on the theoretical distribution vector obtained from $T$.
%\red{Seria transformations aplied to p? en comptes de T? \tp{canviat}}
%
}
\end{figure}

Let $p$ be a distribution arising from a Markov process on the tree $T=12|34$ with transition matrices $M_1$, $M_2$, $M_3$, $M_4$ at the external edges and $M_5$ at the internal edge. The effect of the $T$ leaf-transformations on $p$ is that of replacing some of the external matrices so that both leaves in the same side of the tree share the same matrix. This is achieved in 4 possible patterns, namely $M_1 | M_3$, $M_1 | M_4$, $M_2 | M_3$, or $M_2 | M_4$ (see Figure \ref{transformations}, middle row).
There are four possible ways of computing each of these transformed distributions, and these four ways are not equivalent when applied to a distribution $q$ that has not arisen as a Markov process on $T$. Therefore, in general, we have 16 $T$ leaf-transformations that can be applied to a distribution $q\in \mathbb{R}^{256}$.
%
%When each of these possible ways of computing the $T$-transformations are applied to a distribution vector that does not fit a theoretial distribution on $T$, they will produce $16$ distribution vectors arranged in $4$ groups.

% In practice, the distribution vector $p$ is estimated form the data $\tilde{p}\sim p$, so we deal with a distribution vector which is close but not equal to the theoretical distribution $p$. It turns out that each possible way of computation provides a different vector, and we obtain $16=4 \times 4$ possible $T$ leaf-transformations, which are arranged in groups of 4.

% For each of these possibilities, there are four possible ways of computing the new distribution which provide the same result as far as $p$ fits exactly the distribution $\hat{p}$ corresponding to the original matrices \red{No entenc l'última part de la frase.}.
 %and a complete description on how to compute them in \red{[cita]}.
 %

Analogously, if $T'\neq T$, we also have 16 $T'$ leaf-transformations that act on distributions in $\mathbb{R}^{256}$.
%When they are applied to a theoretical distribution $p'$ that has arisen from $T'$, they result on 4 different distribution vectors which are obtaned by replacing the matrices of $T'$ so that both leaves in the same side of $T'$ share the same matrix;
%but when they are applied to a distribution vector that does not fit a theoretial distribution on $T'$, they produce $16$ distribution vectors arranged in $4$ groups.
%
Remarkably, if $p$ is a distribution that arises from a Markov process on $T$,
these $T'$ leaf-transformations applied to $p$ produce 16 vectors $\hat{p_i}$ that have also arisen on $T$ but with different matrices at the exterior edges (see figure \ref{transformations}, bottom row).
%
% These $T'$ leaf-transformations applied on $p$ produce 16 different vectors $\hat{p}$ that have also arisen on $T$ but with different matrices at the exterior edges.
Moreover, while the former $T$ leaf-transformations applied to $p$ produce \emph{distributions}, the latter may produce non-stochastic vectors since the new parameters may not be stochastic matrices (see Figure \ref{transformations}, for example the transition matrices $M_5^{-1}M_1$ or $M_5^{-1} M_2$ on the 2nd, 3rd and 4th trees depicted in the $13|24$ leaf transformations might not be stochastic). In all these trees, the matrix attached to the interior edge remains untouched and is the matrix $M_5$ of the original tree.

% Both groups of transformations produce distribution vectors arising from a $12|34$ tree, but with different parameters (different transition matrices attached to the edges of the tree).
%

%\vspace{5mm}

\newpage
\subsubsection{The quotient score is well defined}
In this subsection we prove that if $p$ arises from generic parameters on $T=12|34$, the closest positive definite matrix to $F_{13|24}(\hat{p})$ has generically rank strictly larger than four, which is the last theoretical result needed jo justify the basis of \saq method.
%\red{Cal dir algo de perque ho fem per Kimura i no GMM?}
For the sake of simplicity, the proof is carried out on the 3-parameter Kimura model. Then, a deformation argument can be applied to derive that the claim holds for generic Markov matrices.

\newpage
\begin{lema}
	Let $p$ be the theoretical distribution from a $3$-parameter Kimura process on the quartet tree $T=12|34$ with generic transition matrices $M_1$, $M_2$, $M_3$, $M_4$ and $M_5$ (see the tree $T$ of Figure \ref{transformations}). Then, if $\hat{p}$ is obtained by applying a $13|24$ leaf-transformation on $p$, it holds that $$ \delta_4\left(psd\left(F_{13|24}(\hat{p})\right)\right) > 0.$$
\end{lema}

\begin{proof}
	We prove this result for a concrete $13|24$ leaf-transformation on $p$ since the proof for the other transformations is analogous.
	%We assume that  $T$ has $3$-parameters Kimura matrices $M_1$, $M_2$, $M_2$, $M_2$ at the exterior edges and a $3$-parameters Kimura matrix $M_5$ at the interior edge, as in Figure ... Then,
	Namely, we take the $13|24$ leaf-transformation that produces the vector $\hat{p}$ that  corresponds to the tree $T=12|34$ with transition matrix $M_5M_3$ at the edge adjacent to leaf $1$, transition matrix $M_5M_4$ at the edge adjacent to leaf $2$ and transition matrices $M_3$ and $M_4$ at the leaves adjacent to the leaves $3$ and $4$ respectively (see the first tree of the $13|24$ leaf-transformation shown in Figure \ref{transformations}).
	
	It is known that for any matrix $A$, the rank of the nearest positive semidefinite matrix $psd(A)$ is equal to the number of positive eigenvalues of the symmetric matrix $\frac{1}{2}\left(A + A^t\right)$ \citep{nearestPSD}. So, it is enough to show that the symmetric matrix  $B:=\frac{1}{2}\left(F_{13|24}(\hat{p}) + F_{13|24}(\hat{p})^t\right)$ has at least 5 positive eigenvalues.
	To this aim, we use the following known result: if $\Delta_i$ is the $i$-th leading principal minor of a $n \times n$ matrix $A$, the number of sign changes in the sequence $\Delta_0 =1, \Delta_1, \cdots, \Delta_n = det(A)$ equals the number of negative eigenvalues of the matrix $A$ (see \cite{sylvesterCrit} for a proof). Consequently, the number of positive eigenvalues of $A$ equals the number of consecutive leading principal minors $\Delta_i$ without sign changes.
	We will also make use of Sylvester's \emph{law of inertia} which states that any two $n\times n$ symmetric matrices $X$ and $Y$ have the same number of positive, negative and zero eigenvalues if and only if there exists an invertible matrix $H\in\mathbb{R}_{n\times n}$ such that $Y = H^tXH$ (see, for instance, \cite{GMVS} for a precise statement and proof).

	We start by writing the flattening matrix $F_{13|24}(\hat{p})$ in terms of these transition matrices %. The following expression can be deduced from
	\citep{AllmanSemialg}:
	$$F_{13|24}(\hat{p}) = (M_5M_3\otimes M_3)^t\; D \;(M_5M_4\otimes M_4)$$ where $\otimes$ is the Kronecker product of matrices and $D$ is the $16\times 16$ diagonal matrix with the entries of $\dfrac{1}{4}M_5$.
%
% 	According to \cite{nearestPSD}, the rank of the nearest positive semidefinite matrix to any matrix $A$, equals the number of positive eigenvalues of the symmetric matrix $\frac{1}{2}\left(A + A^t\right)$. Since we need to prove that $rk(psd(F_{13|24}(\hat{p}))) > 4$ this is equivalent to proving that the symmetric matrix  $B:=\frac{1}{2}\left(F_{13|24}(\hat{p}) + F_{13|24}(\hat{p})^t\right)$ has at least 5 positive eigenvalues.
%
%	$psd(F_{13|24}(\hat{p}))$ $B:=\frac{1}{2}\left(F_{13|24}(\hat{p}_i) + F_{13|24}(\hat{p}_i)^t\right)$.
% 	We will also use the Sylvester law of inertia (see, for instance, \cite{GMVS} for a precise statement and proof), which states that any two $n\times n$ symmetric matrices $X$ and $Y$ have the same number of positive, negative and zero eigenvalues if and only if there exists an invertible matrix $H\in\mathbb{R}_{n\times n}$ such that $Y = H^tXH$.
%
Consider the symmetric matrix
\footnotesize
\begin{eqnarray*}
H=\left(
\begin{array}{cccc}
1 & 1 & 1 & 1 \\
1 & 1 & -1 & -1 \\
1 & -1 & 1 & -1 \\
1 & -1 & -1 & 1
\end{array}
\right).
\end{eqnarray*}
\normalsize
It is known that any $3$-parameter Kimura matrix $M$ diagonalizes through $H$. Moreover, the multiplicity of 1 as an eigenvalue of $M$ is exactly one if $M$ is generic (see \cite{ARbook}). Write $\{\alpha_k\}_{k=C,G,T}$ for the three eigenvalues of the matrix $M_3$ other than 1, and similarly $\{\beta_k\}_{k=C,G,T}$ and $\{\gamma_k\}_{k=C,G,T}$ for $M_4$ and $M_5$, respectively. We can assume that these eigenvalues are positive since the transition matrices should not be too far from the identity matrix.

Write %$B=F_{13|24}(\hat{p})$ and
$\bar{B} := (H\otimes H)^t B (H\otimes H)$. Because of Sylvester's law of inertia, both matrices, $B$ and $\bar{B}$ have the same number of positive and negative eigenvalues.
The matrix $\bar{B}$ can be written in terms of the eigenvalues of the transition matrices $M_3$, $M_4$ and $M_5$.
%Any $3$-parameter Kimura matrix $M$ diagonalizes with an eigenvalue $m_A$ equal to $1$ and three eigenvalues $m_C$, $m_C$ and $m_C$ strictly smaller than $1$, unless the matrix is a permutation matrix (see \cite{ARbook}). Denote by $\alpha_k$, $\beta_k$ and $\gamma_k$ ( with $k\in\{C,G,T\}$) the tree eigenvalues of $M_3$, $M_4$ and $M_5$ (respectively) different from $1$. We assume that the eigenvalues are all positive since the transition matrices should not be too far from the identity matrix.	
 Then, consider the $5\times 5$ submatrix of $\bar{B}$ generated by removing the last 11 rows and columns of $\bar{B}$:
\begin{eqnarray*}
\footnotesize \bar{B}_5=\frac{1}{4^4}\left(
 \begin{array}{ccccc}
  1 & 0 & 0 & 0 & 0 \\
  0 & \alpha_C \beta_C & 0 & 0 & \alpha_C \beta_C \gamma_C^2\\
  0 & 0 & \alpha_G\beta_G & 0 & 0\\
  0 & 0 & 0 & \alpha_T\beta_T & 0\\
  0 & \alpha_C \beta_C \gamma_C^2 & 0 & 0 & \alpha_C \beta_C \gamma_C^2\\
 \end{array}
 \right),
\end{eqnarray*}
The leading principal minors of $\bar{B}_5$ are equal to the first five leading principal minors of~$\bar{B}$:
\begin{eqnarray*}
& \Delta_1 = 2, \quad \Delta_2 = 4\alpha_C\, \beta_C, \quad \Delta_3 = 8\,\alpha_C\, \alpha_G \beta_C\, \beta_G, \quad \Delta_4 = 16\, \alpha_C\, \alpha_G\, \alpha_T\, \beta_C\, \beta_G\, \beta_T, \\ &
	 \Delta_5 = -32\, \alpha_C^2\, \alpha_G\, \alpha_T\, \beta_C^2\, \beta_G\, \beta_T\, (\gamma_C + 1)\, (\gamma_C - 1)\, \gamma_C^2.
%	$$\small
%	\begin{cases}
%		r_1 = 2 \\
%		r_2 = 4\alpha_C\beta_C \\
%		r_3 = 8\alpha_C \alpha_G \beta_C \beta_G \\
%		r_4 = 16 \alpha_C \alpha_G \alpha_T \beta_C \beta_G \beta_T \\
%		r_5 = -32 (\alpha_C)^2 \alpha_G \alpha_T (\beta_C)^2 \beta_G \beta_T (\gamma_C + 1) (\gamma_C - 1) (\gamma_C)^2
%	\end{cases}
%	$$\textcolor{red}{arreglar coeficients}
%	
\end{eqnarray*}
We have that $\Delta_i>0$  $(i=1,2,3,4)$ because
%The first $4$ leading principal minors of $\bar{B}$ are positive because
$\alpha_k$, $\beta_k$ are positive for any $k\in\{C,G,T\}$. Moreover $\Delta_5$ is also positive since $-(\gamma_C - 1) >0$ for $0<\gamma_C<1$ and $\gamma_G, \gamma_T$ are also positive.
 We conclude that the matrix $\bar{B}$ has at least $5$ positive eigenvalues and so does $B$. This implies that $rk\left(psd\left(F_{13|24}(\hat{p})\right)\right)$ is greater than or equal to $5$ and therefore, $\delta_4\left(psd\left(F_{13|24}(\hat{p}_i)\right)\right) > 0$.
\end{proof}

% \begin{figure}[h]
% \begin{center}
% \includegraphics[scale=0.5]{heterogenoustree_kolaczkowski.eps}
% \end{center}
% %\vspace{3mm}
%\caption{\label{kolaczkowski} \footnotesize
%Heterogenous data taken from \cite{Kolaczkowski2004}: two categories of the same size are considered, both evolving under the GM model on the two trees depicted above with the branch lenghts indicated. The internal branch length takes the same value
%in both categories and varies from 0.01 to 0.4 in steps of 0.05.
%}
%\end{figure}

% ternary plot GTR
\subsection{Random branch lengths}

Here we present the ternary plots of the performance of \saq when applied to the alignments generated on the tree $12|34$ under the GM and homGTR models, with length 1~000 bp and 10~000 bp and random branch lengths uniformly distributed between 0 and~3.

\begin{figure}[h]
 \begin{center}
 \includegraphics[scale=0.15]{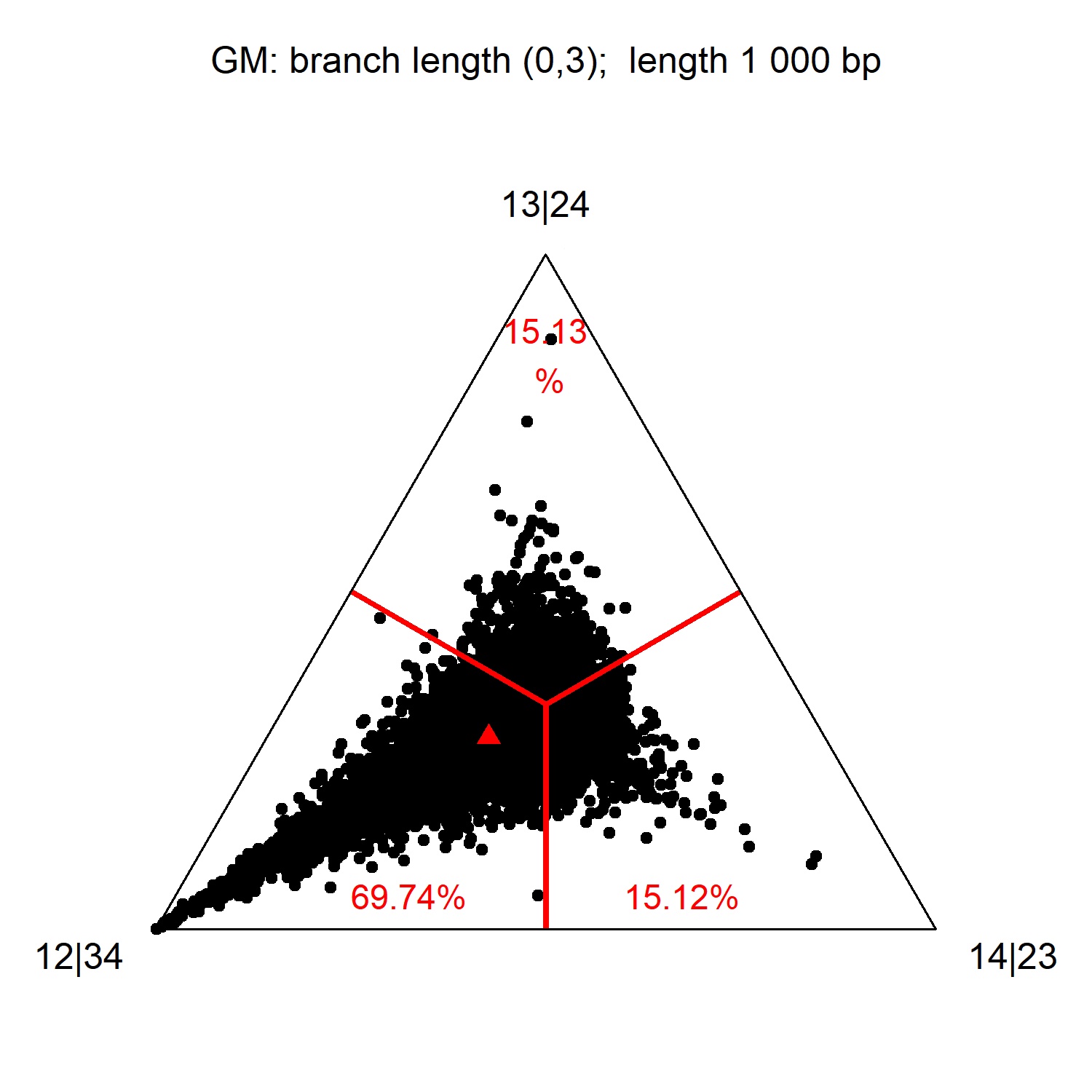}
 \includegraphics[scale=0.15]{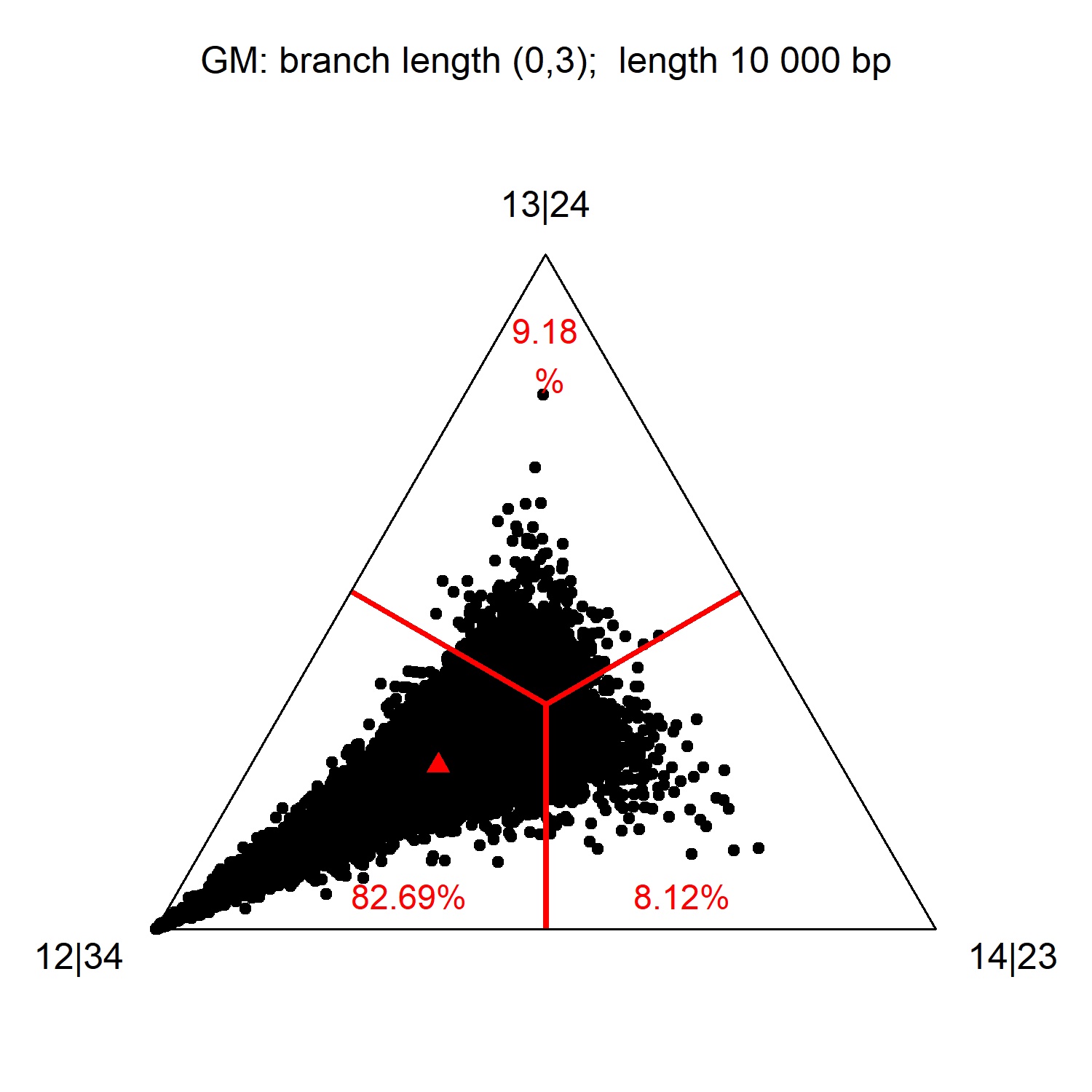}\\
 \includegraphics[scale=0.15]{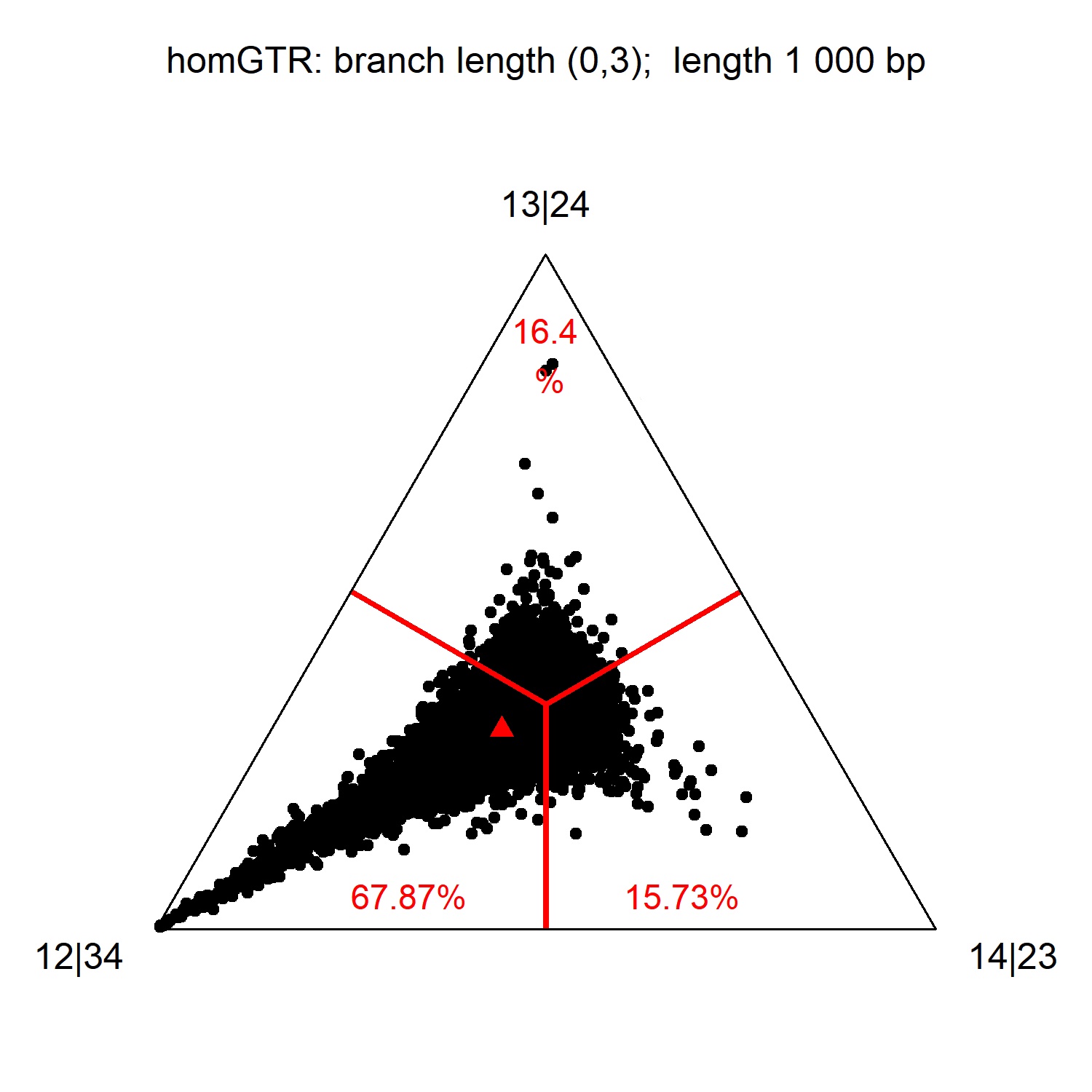}
 \includegraphics[scale=0.15]{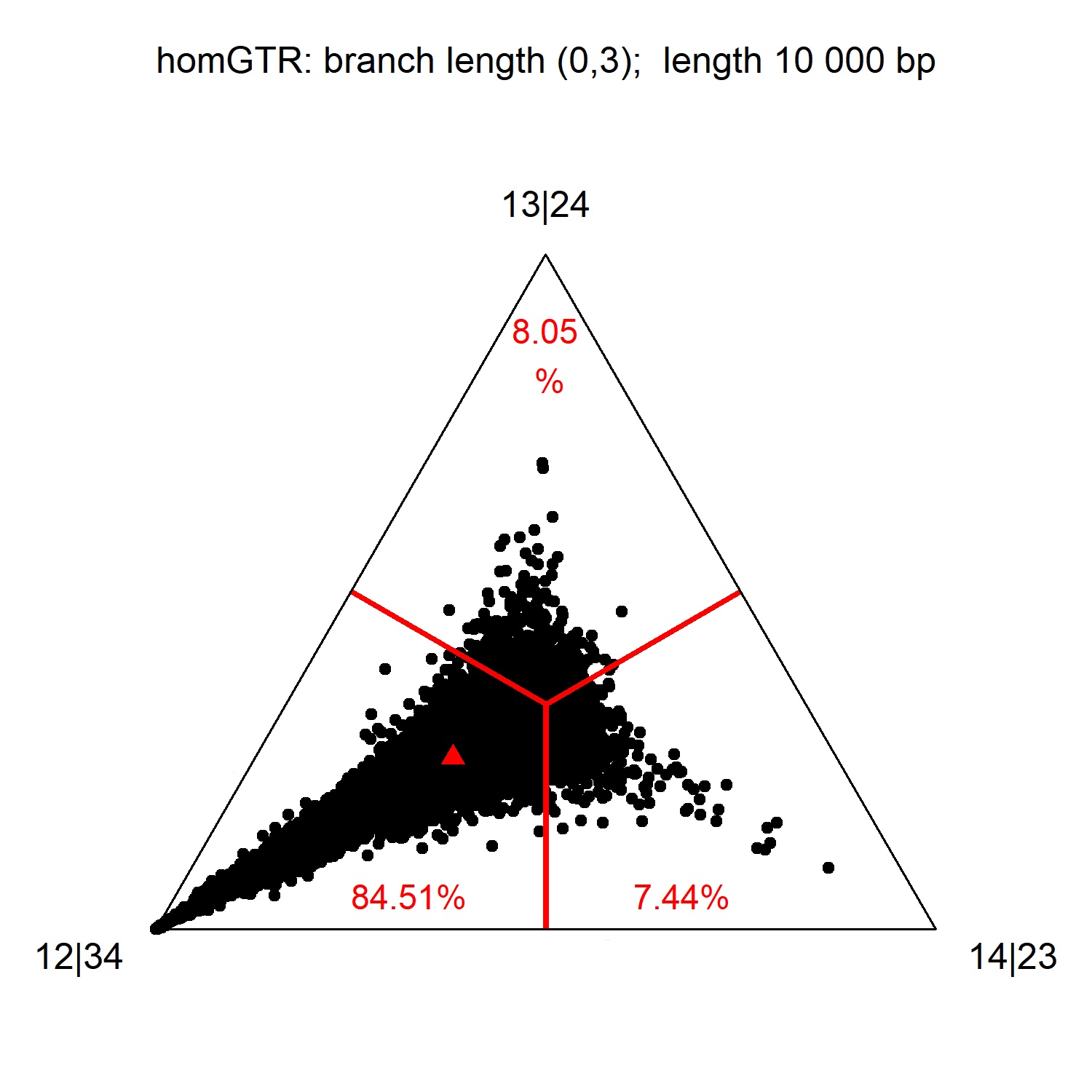}
 \end{center}
 %\vspace{3mm}
\caption{\label{ternary_03} \footnotesize
%\tp{PENDENT: editar figures}
Ternary plots corresponding to the weights of \saq applied to 10000 alignments
generated under the $12|34$ tree with random branch lengths uniformly distributed between
0 and 3. On each triangle the bottom-left vertex represents the underlying true topology
$12|34$, the bottom-right vertex is topology $13|24$ and the top vertex is $14|23$.
Each triangle is divided into three regions according to which tree is selected by the method. The figures in red represent  the percentage of alignments that correspond to the corresponding region according to \saq.
Top: correspond to data generated under GMM; bottom: data
generated under homogenenous GTR. Left : 1000 bp; Right: 10000 bp.).}
\end{figure}

\end{document}